\documentclass[a4paper,11pt]{article}
\pdfoutput=1

\usepackage[noblocks]{authblk}
\usepackage{here}
\usepackage{subfig}
\usepackage[pdftex]{graphicx}
\usepackage{color}
\usepackage{braket}
\usepackage{amssymb, amsmath}
\usepackage{feynmf}
\usepackage{bm}
\usepackage{url}
\usepackage{slashed}
\usepackage{caption}

\usepackage{jheppub}
\usepackage[compat=1.1.0]{tikz-feynhand}
\usepackage{hyperref}
\usepackage[samesize]{cancel}
\usepackage{fancyhdr}

\newcommand{\be}{\begin{eqnarray}}
\newcommand{\ee}{\end{eqnarray}}

\newcommand{\beq}{\begin{equation}}
\newcommand{\eeq}{\end{equation}}
\newcommand{\beqa}{\begin{eqnarray}}
\newcommand{\eeqa}{\end{eqnarray}}

\newcommand{\lmk}{\left(}
\newcommand{\rmk}{\right)}
\newcommand{\lkk}{\left[}
\newcommand{\rkk}{\right]}
\newcommand{\lnk}{\left\{}
\newcommand{\rnk}{\right\}}

\begin{document}

\title{\boldmath On the inefficiency of fermion level-crossing under the parity-violating spin-2 gravitational field}
\author[a,b,c]{Kohei Kamada}
\author[d,e,c]{and Jun'ya Kume}
\affiliation[a]{School of Fundamental Physics and Mathematical Sciences, Hangzhou Institute for Advanced Study,University of Chinese Academy of Sciences (HIAS-UCAS), Hangzhou 310024, China}
\affiliation[b]{International Centre for Theoretical Physics Asia-Pacific (ICTP-AP), Beijing/Hangzhou, China}
\affiliation[c]{Research Center for the Early Universe (RESCEU), Graduate School of Science, The University of Tokyo, Hongo 7-3-1
Bunkyo-ku, Tokyo 113-0033, Japan}
\affiliation[d]{Dipartimento di Fisica e Astronomia ``G. Galilei'', Universit\`a degli Studi di Padova, via Marzolo 8, I-35131 Padova, Italy}
\affiliation[e]{INFN, Sezione di Padova, via Marzolo 8, I-35131 Padova, Italy}

\emailAdd{kohei.kamada@ucas.ac.cn}
\emailAdd{junya.kume@unipd.it}

\subheader{{\rm RESCEU-9/24}}

\abstract{Gravitational chiral anomaly connects the topological charge of spacetime and the chirality of fermions.
It has been known that the chirality is carried by the particles (or the excited states) and also by vacuum. While the gravitational anomaly equation has been applied to cosmology, distinction between these two contributions has been rarely discussed. 
In the study of gravitational leptogenesis, for example, lepton asymmetry associated with the chiral gravitational waves (GWs) sourced during inflation is evaluated only by integrating the anomaly equation.
This approach, however, does not reveal how these two contributions are distributed in this scenario.
Meanwhile, the dominance of vacuum contribution is observed in some specific types of Bianchi spacetime with parity-violating gravitational fields.
One may wonder whether such a vacuum dominance takes place also in the system with chiral GWs around the flat background, which is more suitable for application to realistic cosmology.
In this work, we apply an analogy between U(1) electromagnetism and the weak gravity to the spacetime that captures the characteristics of
the one considered in the gravitational leptogenesis.
With this approach, we try to obtain intuitive understanding of the fermion chirality generation under the parity-violating spin-2 gravitational field.
By assuming the emergence of Landau level-like dispersion relation in our setup, we observe that spin-2 nature seems to make the level-crossing inefficient, indicating that the chrial charge is likely to accumulate in the vacuum. 
On this basis, phenomenological implications for gravitational leptogenesis are discussed.
}

\maketitle
\flushbottom

\section{Introduction}
In the relativistic field theory, which successfully describes particles and forces in nature, classical conservation law is sometimes violated once the fields are quantized. Such a violation of the classical symmetry at the quantum level is called quantum anomaly.
One of the most well known examples is the chiral anomaly in a massless quantum electrodynamics (QED), first discovered by S.~Adler, J.~Bell and R.~Jackiw~\cite{Adler:1969gk,Bell:1969ts}. 
There, the chiral current of a Dirac fermion is connected to the topological current of the external U(1) gauge field.
Such an anomalous violation of the axial symmetry, or the chiral anomaly, has a variety of implications to the phenomenology.
For example, anomalous violation of $B + L$, the sum of baryon number $B$ and lepton number $L$ in the Standard Model (SM) of particle physics, is derived as a net contribution of the chiral anomaly for the SU(2)$_W$ as well as U(1)$_Y$ gauge interaction~\cite{tHooft:1976rip}. 
One of the most important consequence of this $B + L$ anomaly is the electroweak sphaleron process~\cite{Manton:1983nd,Klinkhamer:1984di,Kuzmin:1985mm}, which is essential for the electroweak baryogenesis~\cite{Kuzmin:1985mm} as well as the leptogenesis~\cite{Fukugita:1986hr} and realizes the non-trivial charge (re)distribution in the equilibrium plasma in the early Universe~\cite{Khlebnikov:1988sr, Harvey:1990qw,Khlebnikov:1996vj}.
Also it is recently noticed that through the chiral anomaly the helical 
U(1)$_Y$ hypermagnetic field generation during axion 
inflation leads to the simultaneous generation of $B$+$L$ asymmetry~\cite{Domcke:2018eki,Domcke:2019mnd,Domcke:2022kfs}, 
together with the $B$+$L$ generation from the hypermagnetic helicity 
decay at the electroweak symmetry breaking~\cite{Giovannini:1997eg,Kamada:2016cnb}, 
which can be responsible for the present baryon asymmetry
of the Universe.

For the U(1) gauge theory, the chiral anomaly equation indicates that the aligned electric field and magnetic field produces the chiral charge of fermion. This process is elegantly described with 
the Landau levels due to the magnetic field.
It has been shown that only the lowest Landau level (LLL), which smoothly connects the positive frequency and negative frequency mode, participates in this parity-violating process while higher Landau level preserving parity does not~\cite{Nielsen:1983rb, Domcke:2018eki}. 
The excitation in the LLL caused by the electromagnetic field yields the chiral charge consistent with the prediction of chiral anomaly equation.
One should, however, note that chirality connected to the gauge field topology in anomaly equation consists of the asymmetry both in the excitation and the vacuum.
The later is known as the eta invariant~\cite{Atiyah:1963zz,Atiyah:1968mp}, which can be evaluated with the dispersion of the Dirac field.
As discussed in Ref.~\cite{Domcke:2018gfr}, substantial chirality is actually accumulated in the case of chirality production due to the homogeneous SU(2) gauge field. 
Note that the vacuum contribution also exists in the induced SU(2) current, which was shown to be renormalized with the running coupling constant.
This observation indicates that when applying the prediction of chiral anomaly to the phenomenology, proper distinction between the contribution from excitation and vacuum might be required.

While those mentioned above are for the gauge fields, one can also consider the gravitational contribution to the chiral anomaly~\cite{Kimura:1969iwz,Delbourgo:1972xb,Eguchi:1976db,AlvarezGaume:1983ig} and its phenomenological consequences.
For example, the topological charge of the background metric and simultaneously chiral charge of fermions become non-vanishing when the circular polarization of gravitational waves (GWs) are generated or growing as in the models of axion inflation.
For the SM plasma, if the right-handed neutrino is decoupled from the theory up to the inflationary scale, lepton number can be generated in such axion inflation models~\cite{Alexander:2004us}. 
This scenario, which may be responsible for the observed baryon asymmetry of the Universe, is called as gravitational leptogenesis and investigated in the field of particle cosmology~\cite{Alexander:2004us,Lyth:2005jf,Fischler:2007tj,Maleknejad:2012wqk,Maleknejad:2014wsa,Kawai:2017kqt,Caldwell:2017chz,Papageorgiou:2017yup,Adshead:2017znw,Kamada:2019ewe,Kamada:2020jaf}. 

In the literature, however, no distinction between the excitation and vacuum contribution has been drawn in the evaluation of lepton number.
The conversion of ``vacuum lepton charges'' to the observable baryon charges would not be described by the usual kinetic equation with the electroweak sphaleron process unlike those in the thermal plasma, and, in the worst case, they might not be converted at all.
A conservative prediction of the net baryon asymmetry would be those by counting the contribution from excitation.
The question then becomes, in the gravitational leptogenesis which is more dominant in the system, the excitation or the vacuum contribution?
In fact, dominance of the chirality accumulated in vacuum is also observed in the gravitational systems. One example is the Bianchi type-II spacetime recently studied in Ref.~\cite{Stone:2023qln}. Another example is the Bianchi type-IX spacetime discussed in the old seminal papers~\cite{Gibbons:1979kq,Gibbons:1979ks} by Gibbons.
The later one is of our interest since the spacetime can be decomposed into the closed Friedmann-Lemaitre- Roberson-Walker (FLRW) universe and the standing chiral GW~\cite{Grishchuk:1974ny,King:1991jd}.

In this work, we investigate the effect of parity-violating weak spin-2 gravitational field on the massless Dirac field, which cannot directly be extrapolated from the results of Bianchi spacetimes.
To this end, we find a ``simple'' but reasonable configuration of the metric 
that captures well the characteristics of chiral GWs assumed in the gravitational leptogenesis
and is similar to the homogeneous electromagnetic field.
However, it turns out to be difficult to find an analytical solution that can be interpreted in a physical sense.
Compared to the solvable Bianchi type-IX case, this is partly due to the absence of a globally defined momentum. 
Instead of trying to find solutions of the Dirac equation, we make use of an analogy between the classical electromagnetism and the weak gravitational field~\cite{Mashhoon:2003ax,FilipeCosta:2006fz}. We then discuss whether fermion chirality can be excited in the gravitational leptogenesis.
In order to make this inference based on the analogy more convincing, we also clarify how the chirality generation of fermions in the Bianchi type-IX spacetime is analogous to the SU(2) gauge field case.

The rest of papers is organized as follows.
In Sec.~\ref{sec:GL_review}, we briefly review the gravitational leptogenesis scenario and the conventional evaluation of lepton number produced by inflationary chiral GWs. 
In Sec.~\ref{sec:gauge}, we give a review on the chirality production of fermions under the gauge fields. This provides the basis to understand the physics of chirality generation in the gravitational systems. We discuss the analogy between cases of the Bianchi type-IX spacetime  and SU(2) gauge field in Sec.~\ref{sec:bianchi}. In Sec.~\ref{sec:weak_spin2}, we introduce the metric configuration that is similar to the homogeneous electromagnetic field with non-vanishing topological charge and captures the characiterstics of those generated in the gravitational leptogenesis scenario.
We finally make a comment on the implication to the gravitational leptogenesis in Sec.~\ref{sec:discussion}.
We use the notations following Refs.~\cite{Alexander:2009tp,Jackiw:2003pm}:
\begin{itemize}
    \item sign of the metric: $g_{\mu\nu} = (- + + +)$.
    \item Greek indices ($\mu, \nu, ...$) run 0 to 3, Latin indices ($i, j, ...
    $) run 1 to 3.
    \item Levi-Chivita symbol: $\epsilon^{0123} = 1$.
    \item Chern-Pontryagin density: $R\tilde{R} \equiv -\frac{1}{2}\frac{\epsilon^{\alpha\beta\gamma\delta}}{\sqrt{-g}}R_{\alpha\beta\rho\sigma}R_{\gamma\delta}^{\ \ \rho\sigma}$.
    \item $\gamma$ matrices in Weyl rep.:
    \begin{equation*}
        \gamma^{\mu} = \left(
    \begin{array}{cc}
    0 & \sigma^{\mu}\\
    \bar{\sigma}^{\mu} & 0
    \end{array}
    \right)
    = \left\{
    \left(
    \begin{array}{cc}
    0 & I_{2\times2}\\
    I_{2\times2} & 0
    \end{array}
    \right),
    \left(
    \begin{array}{cc}
    0 & \sigma_i\\
    -\sigma_i & 0
    \end{array}
    \right)
    \right\},
    \end{equation*}
    where $\sigma_i$ are the Pauli matrices.
\end{itemize}

\section{Gravitational leptogenesis}\label{sec:GL_review}
In this section, we briefly review the gravitational leptogenesis scenario, which is the starting point of our study. In the SM, due to the absence of the right-handed neutrinos, the lepton number is gravitationally violated through the chiral anomaly as~\cite{Kimura:1969iwz,Delbourgo:1972xb,Eguchi:1976db,AlvarezGaume:1983ig}
\beq
\nabla_{\mu} J_L^{\mu} = \frac{3}{384\pi^2}R\tilde{R},\label{eq:lepton_ganom_SM}
\eeq 
where $J^\mu_L$ is the lepton current and $R\tilde{R}$ is the gravitational Chern-Pontryagin density. The coefficient $3$ comes from the number of family of leptons.
This gravitational anomaly equation~\eqref{eq:lepton_ganom_SM} indicates that lepton number should be generated in the spacetime with non-vanishing $R\tilde{R}$. 
Among such spacetime configurations, parity-violating spin-2 perturbations (or circularly polarized GWs) around flat Friedmann-Lemaitre-Roberson-Walker (FLRW) spacetime has been discussed in the context of inflationary phenomenology. More explicitly, the spacetime metric is expressed as
\begin{align}
\mathrm{d}s^2 = a^2(\eta) [-\mathrm{d}\eta^2+(\delta_{ij} + h_{ij}(\eta, {\bm x}))\mathrm{d}x^i \mathrm{d}x^j],\label{eq:inf_tensor_pert} 
\end{align}
where $|h_{ij}| \ll 1$ is the spin-2 perturbation around the conformally flat spacetime. 
This perturbation is known to propagate as GW expressed with two independent polarization mode and satisfies transverse-traceless (TT) condition
\beq
h^i_{\ i} = 0,\ \ \partial^j h_{ij} = 0, \label{eq:TTgauge}
\eeq
where the indices are raised and lowered by $\eta_{\mu\nu} = \mathrm{diag} (-1,1,1,1)$.
Similarly to the electromagnetic waves, one can define the circular polarization basis (left and right) for GWs.
Interestingly, circularly polarized GWs can be generated and continuously contribute to the expectation value of the Chern-Pontryagin density $\langle R {\tilde R} \rangle$ in some models of axion inflation.
For example, in the model with gravitational Chern-Simons coupling of inflaton $\phi R {\tilde R}$~\cite{Alexander:2004us,Lyth:2005jf}, which is regarded as the minimal model for the gravitational leptogenesis, $\langle R {\tilde R} \rangle$ becomes constant during inflation~\cite{Fischler:2007tj,Kamada:2020jaf} (see also Refs.~\cite{Maleknejad:2012wqk,Maleknejad:2014wsa,Kawai:2017kqt,Caldwell:2017chz,Papageorgiou:2017yup,Adshead:2017znw} for the other realizations).
According to the anomaly equation, lepton numbers  would be simultaneously produced in the early universe, which could explain the observed baryon number excess~\cite{Alexander:2004us,Lyth:2005jf} through the conversion by the electroweak sphaleron process~\cite{Klinkhamer:1984di,Kuzmin:1985mm,Fukugita:1986hr}. This lepton number production associated with the circularly polarized GW generation is called gravitational leptogenesis. 

Conventionally, the lepton number produced during inflation is evaluated by integrating the spatially averaged anomaly equation up to the linear order in $h$:
\begin{align}
    &\langle n_L(\eta_f) \rangle = a(\eta_f) J^0_L(\eta_f) \notag \\
    &\quad 
    = \frac{3}{192\pi^2a(\eta_f)^3}\int_{\frac{k}{a(\eta_f)}<\Lambda} \frac{\mathrm{d}^3k}{(2\pi)^3} {k\left[\left(|(h^\mathrm{R}_{{\bf k}}(\eta_\mathrm{f}))'|^2-k^2|h^\mathrm{R}_{{\bf k}}(\eta_\mathrm{f})|^2\right) - \left({\rm R} \leftrightarrow {\rm L}\right) \right]},\label{eq:GL_lepton}
\end{align}
where $\eta_f$ is conformal time at the end of inflation and $h^\mathrm{R/L}_{\bf k}(\eta)$ is the mode function of quantized perturbation (graviton)
in the circular polarization basis (see App.~\ref{app:GL_min_config} for more details). 
Here in the integration, we introduce the cutoff scale $\Lambda$, which yields a factor $(\Lambda/H)^4$ with $H$ being the inflationary Hubble scale. 
Note that no definitive conclusion has been reached on what scale should be taken as this $\Lambda$. For example, $\Lambda$ may be taken as the Hubble scale during inflation after proper renormalization procedure~\cite{Fischler:2007tj,Kamada:2020jaf}, while it could be the scale of a UV physics such as the mass of the right-handed Majorana neutrino~\cite{Alexander:2004us}. Therefore, here we shall not go into the details of the evaluation of Eq.~\eqref{eq:GL_lepton}.

As in the vanilla leptogenesis scenario~\cite{Fukugita:1986hr}, the lepton number~\eqref{eq:GL_lepton} is assumed to be converted to the baryon number in the thermal equilibrium as~\cite{Harvey:1990qw,Khlebnikov:1996vj}
\begin{align}
    n_B|_{\rm eq} = - \frac{28}{79}n_L(\eta_f).\label{eq:sphaleron}
\end{align}
Therefore, if sufficiently large asymmetry is produced in the circular polarization of inflationary GWs, the observed value of $n_B/s$ can be explained with this anomalously generated lepton number\footnote{Even if the lepton asymmetry at the end of inflation is significantly small, $n_L/s$ might be larger by assuming the inefficient reheating process~\cite{Kamada:2019ewe}}.

However, this evaluation never clarifies how the chirality of fermion is generated and to the best of our knowledge, there has been no investigation of the fermion field equations under the chiral spin-2 background assumed in the gravitational leptogenesis. 
As we will see below, the fermion chirality generally consists of two different contributions and its distribution is in fact significantly differs depending on the external field sourcing the chirality.
Before discussing the chirality generation under the gravitational fields, we review those under the gauge fields in the flat 3+1 spacetime, which provides a basis to investigate the former in the context of gravitational leptogenesis.

\section{Chirality production of fermions under helical gauge fields}\label{sec:gauge}
In this section, we review the chirality production of fermions under external gauge fields in terms of the level crossing. 
Throughout this section, we consider flat 3+1 spacetime.
While the cosmological expansion is turned off, this provides a basis of understanding particle production in the scenarios of our interest.
For a U(1) gauge field, the physics of chirality production can be nicely understood with the creation of Landau levels. While it accounts for all the chirality in the U(1) case, vacuum contribution becomes dominant in the case of non-Abelian gauge field. While the latter example provide the understanding on when the contribution of vacuum becomes important, the former one is expected to be useful for understanding the effect of weak gravitational field as we will see later.

\subsection{Chiral anomaly and Landau levels in U(1) theory}\label{sec:u1}
Let us start with massless Dirac fermions charged under the U(1) gauge field~\cite{Nielsen:1983rb}.
The Lagrangian is given by
\begin{equation}
{\cal L} = \mathrm{i} {\bar \Psi} \gamma^\mu (\partial_\mu + \mathrm{i} e A_\mu) \Psi, 
\end{equation}
where $\Psi = \left( \begin{array}{c} \psi_L \\ \psi_R \end{array}\right)$ is the Dirac fermion ($\psi_{R/L}$ are the right- and left-handed fermions, respectively), $e$ is its U(1) charge, and $A_\mu$ is the U(1) gauge field. 
We define the
field strength tensor and its dual 
as
\begin{align}
    F_{\mu\nu} = \partial_{\mu}A_{\nu}-\partial_{\nu}A_{\mu}\label{eq:FS},\\
    \tilde{F}^{\mu\nu} = \frac{1}{2}\epsilon^{\mu\nu\alpha\beta}F_{\alpha\beta}.
\end{align}
The electric field and the magnetic field are written as
\begin{align}
E_i = F_{i0},\\    
B^i = \frac{1}{2}\epsilon^{0ijk}F_{jk}.
\end{align}
The Chern-Pontryagin density is expressed in the following way,
\beq
F_{\mu\nu} \tilde{F}^{\mu \nu } \equiv \frac{1}{2}\epsilon^{\mu\nu\alpha\beta}F_{\mu\nu}F_{\alpha\beta} = -4\vec{E}\cdot \vec{B}.\label{eq:FFdual}
\eeq
We immediately find that the parallel $\vec{E}$ and $\vec{B}$ fields give non-zero $F\tilde{F}$. 
In the following we investigate the particle production 
in the presence of non-vanishing background Chern-Pontryagin density.

For concreteness, we consider the following background 
gauge field configuration,
\begin{equation}
A_{\mu} = (0,0,Bx,-Et), \label{U1homoconfig}
\end{equation}
which gives a homogeneous electric and magnetic field in the $z$ direction, $E_z = E, B_z = B$, and hence $F_{\mu\nu} \tilde{F}^{\mu \nu } = - 4 EB$~\cite{Domcke:2018eki}.
First we turn on only the magnetic field. 
Then, the quantization in the $x$-$y$ plane gives us the 
energy dispersion relation in the momentum space, known as the Landau levels, 
\begin{equation}
\omega_n^2 = \Pi_z^2 + (2n +1) e B -2 e B S_z, n=0,1,2, \cdots.  
\end{equation}
Here $\omega_n$ is the energy, $\Pi_z=p_z-e A_z (=p_z \ (\text{for} \ A_z=0))$ is the canonical conjugate momentum in the $z$ direction, the integer $n$ represents the quantized energy levels in the $x$-$y$ plane, and $S_z=\pm1/2$ is the spin 
of the fermion in the $z$ direction. 
The LLL corresponds to the state with $n=0$ and $S_z=1/2$, that is, 
$\omega_n^2 = \Pi_z^2$ or $\omega_n = \pm \Pi_z$.
+ and - correspond to the right- and left-handed fermions, respectively~\cite{Nielsen:1983rb,Landsteiner:2016led}. 
Higher Landau levels (HLLs) correspond to $\omega_n = \pm \sqrt{\Pi_z^2 + 2 n e B} \ (n>0)$, which are gapped and parity-symmetric. 
The vacuum of this system is the Dirac sea, 
where the negative energy states in the HLL, $\omega_n = -\sqrt{\Pi_z^2 + 2 n e B}$, and those in the LLL, are fully occupied. 

\begin{figure}[htbp]
\begin{center}
\includegraphics[width=0.6\columnwidth]{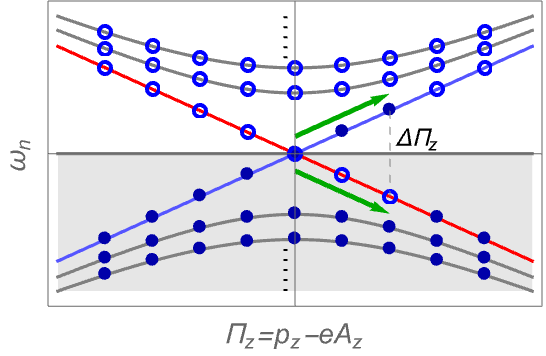}
\caption{\small
The Landau levels for massless Dirac fermions.
Blue and red lines represent the LLL for the right- and left-handed fermions, respectively, and 
gray curves represent HLLs. 
Filled (empty) circles represent the occupied (empty) states, and gray shaded region means the Dirac sea.
After turning 
on an electric field adiabatically, right-handed fermions in the LLL
obtain the positive energy while left-handed fermions 
in the LLL develop the holes. The states in the HLLs look unchanged. Note that the state is continuously distributed in the present case. }
\label{fig:U1LL}
\end{center}
\end{figure}

Now we turn on the electric field adiabatically in the $z$ direction
to the system in the vacuum. 
The classical equation of motion for the fermion is $\partial \Pi_z/\partial t = e E$, which means that each particle in a energy level acquires a 
momentum
\begin{equation}
    \Delta \Pi_z = e \int E \mathrm{d}t, 
\end{equation}
along with the corresponding Landau levels. 
While the particles in the HLL stay in the Dirac sea, 
the right-handed fermions in the LLL are shifted to the positive
energy state and the left-handed fermions in the LLL develop the hole
as shown in Fig.~\ref{fig:U1LL}, 
which corresponds to the level crossing from the shift of the canonical momentum in the presence of the electric field. 
As a result, a chiral asymmetry is induced in the system. 
Quantitatively, the number of induced right- and left-handed fermions are evaluated as
\begin{align}
\Delta Q_{\rm R} &= \int \frac{eE}{2\pi} \mathrm{d}t \mathrm{d}z \int \frac{eB}{2\pi} \mathrm{d}x \mathrm{d}y = \frac{e^2}{4 \pi^2} \int \mathrm{d}^4 x EB, \\
\Delta Q_{\rm L} &= -\int \frac{eE}{2\pi} \mathrm{d}t \mathrm{d}z \int \frac{eB}{2\pi} \mathrm{d}x \mathrm{d}y = - \frac{e^2}{4 \pi^2} \int \mathrm{d}^4 x EB, 
\end{align}
where we have taken into account the quantization condition of the momentum in the $z$ direction, $\int \Pi_n/2\pi \mathrm{d}z \in \mathbb{Z}$, and the Landau degeneracy per unit area in the $x$-$y$ plane, $eB/2\pi$. 
Consequently, we obtain the variation of the axial charge $Q_5 \equiv Q_{\rm R}-Q_{\rm L}$ as
\begin{equation}
    \Delta Q_5 = \Delta Q_{\rm R}-\Delta Q_{\rm L} = \frac{e^2}{2 \pi^2} \int \mathrm{d}^4 x EB. 
\end{equation}
We find that Lorentz-covariant version of this expression 
is nothing but the chiral anomaly equation (in the covariant definition~\cite{Landsteiner:2016led})
\begin{equation}
    \partial_\mu j^\mu_5 = \frac{e^2}{2 \pi^2} EB = - \frac{e^2}{8\pi^2} F_{\mu\nu} \tilde{F}^{\mu \nu }, 
\end{equation}
where the axial current $j^\mu_5 = (Q_5, j^i_5)$ 
is defined as $j^\mu_5 \equiv {\bar \Psi} \gamma^\mu \gamma^5 \Psi$. 
This description clearly shows that the Landau level crossing is a powerful tool to examine the particle production as an excitation 
in the chiral anomaly. 
Moreover, this particle production with the field configuration~\eqref{U1homoconfig} is used to 
investigate the induced current during axion inflation~\cite{Domcke:2018eki} (see also Refs.~\cite{Gorbar:2021rlt,Fujita:2022fwc}), 
where the gauge fields are amplified at the Hubble horizon scale 
and can be taken as (random) homogeneous field at smaller scales.

\subsection{Non-Abelian gauge theory and eta-invariant}\label{sec:gauge_su2}
We next review the chirality production
under external SU(2) gauge field, $A^a_\mu$, 
discussed in Ref.~\cite{Domcke:2018gfr}. 
We consider the following Lagrangian,
\begin{equation}
{\cal L} = \mathrm{i} {\bar \Psi} \gamma^\mu (\partial_\mu - \mathrm{i} A_\mu^a T^a) \Psi, 
\end{equation}
where $T^a=\sigma^a/2$ is the generator of the SU(2) gauge group.
The field strength tensor and its dual are given as
\begin{align}
    F_{\mu\nu}^a &= \partial_{\mu}A_{\nu}^a-\partial_{\nu}A_{\mu}^a + \epsilon^{0abc} A^b_\mu A^c_\nu \label{eq:FSsu2},\\
    \tilde{F}^{a \mu\nu} &= \frac{1}{2}\epsilon^{\mu\nu\alpha\beta}F_{\alpha\beta}^a.
\end{align}
We may define the electric and magnetic fields for the SU(2) 
field as
\begin{align}
E_i^a &= F_{i0}^a,\\    
B^{a i} &= \frac{1}{2}\epsilon^{0ijk}F_{jk}^a, 
\end{align}
such that the Chern-Pontryagin density is written as 
\begin{align}
    F_{\mu\nu}^a \tilde{F}^{a\mu\nu} = -4E^{a}_iB^{ai}.
\end{align}

One could find the exactly same physics for chirality production as the case of U(1) gauge theory supposing a background field configuration,
$A_{\mu}^a = \bar{A}_{\mu}n^a$ with $\bar{A}_{\mu} = (0, 0, -Bx, Et)$. 
Here $n^a$ is an arbitrary constant unit vector which projects the SU(2) gauge group onto its U(1) subgroup.
Once we have in mind the gauge field amplification during axion inflation through the Chern-Simons coupling, $(\phi/4 \Lambda) \mathrm{Tr} F_{\mu\nu}^a \tilde{F}^{a\mu\nu}$ ($\phi$ is the inflaton and $\Lambda$ is the mass scale related to axion decay constant), however, this configuration is turned out to be
unstable for large $(\partial \phi/\partial t)/\Lambda$ due to the non-linear term in $F_{\mu\nu}^a$~\cite{Domcke:2018rvv}. 
Instead, it has been found that there is a 
homogeneous and isotropic attractor solution 
in the presence of the homogeneous axion dynamics as 

\begin{align}
    A_0^a = 0, \quad A_i^a = -f(t)\delta_i^a, \label{configCN}
\end{align}
where $f(t)$ is determined by the homogeneous axion velocity $(\partial \phi/\partial t)$. 
Such a field configuration has been extensively studied in the context of the chromo-natural inflation~\cite{Adshead:2012kp} (see also Ref.~\cite{Maleknejad:2012wqk}). 
The electric and magnetic field, as well as the Chern-Pontryagin density for this field configuration is given as
\begin{align}
    E^a_i &= {\dot f} \delta^a_i \\
    B^{a i} &=  f^2 \delta^{ai} \\
    F_{\mu\nu}^a \tilde{F}^{a\mu\nu} &= -4E^{a}_{i}B^{ai} = -12\dot{f}f^2 = -4{(f^3)}^\cdot.
\end{align}
Let us stress that here a non-vanishing magnetic field 
is provided by the homogeneous vector potential, which is distinct from the Abelian case.
In the following, we consider the field configuration 
given by Eq.~\eqref{configCN} and investigate 
the particle production from this background, changing $f$ adiabatically.

One can explicitly solve the Dirac equation in the momentum space under this SU(2) field configuration \eqref{configCN} with $f$ being constant:
\begin{align}
    \left[\mathrm{i}\partial_t \pm ({\bm \sigma}\cdot{\bm p} + f {\bm \sigma}\cdot{\bm T})\right] \psi_{\rm L/R}(t, {\bm p}) = 0,
\end{align}
where $+ (-)$ refers left- (right-)handed fermions, respectively, and the SU(2) generators $T^a$ acts on the gauge indices of the Dirac field while the Pauli matrices ${\sigma}^i$ acts on its spin indices.
Without loss of generality, one can take the momentum along the z-direction, ${\bm p} = p\hat{e}_z$, and introduce the following eigenbases of the spin and gauge degrees of freedom $\chi_{\bm p}^{\pm}$ and $t_{\bm p}^{\pm}$ satisfying
\begin{align}
    &\lmk \hat{\bm p} \cdot {\bm \sigma}\rmk\chi_{\bm p}^{\pm} = \pm \chi_{\bm p}^{\pm}, \quad \lmk \hat{\bm p} \cdot {\bm T}\rmk t_{\bm p}^{\pm} = \pm \frac{1}{2}t_{\bm p}^{\pm}, \\
    &\sigma_\pm \chi_{\bm p}^{\pm} = 0,  \quad \sigma_\mp \chi_{\bm p}^{\pm} = 2 \chi_{\bm p}^{\mp}, \quad T_\pm t_{\bm p}^{\pm}=0, \quad T_\mp t_{\bm p}^{\pm}=t_{\bm p}^{\mp},
\end{align}
where $\sigma_\pm \equiv \sigma_1 \pm \mathrm{i} \sigma_2$ and $T_\pm \equiv T_1 \pm \mathrm{i} T_2$ are ladder operators. Then, 
the Dirac field can be expanded with these bases
\begin{align}
    \psi_{\rm L/R}(t,{\bm p}) = \sum_{s,m = \pm} \psi^{(s,m)}_{\rm L/R}(t,{\bm p})\chi_{\bm p}^{s}t_{\bm p}^{m},
\end{align}
and noting that ${\bm \sigma}\cdot{\bm T}= (\sigma_+ T_- + \sigma_- T_+)/2 + \lmk \hat{\bm p} \cdot {\bm \sigma}\rmk \lmk \hat{\bm p} \cdot {\bm T}\rmk$ the Dirac equation finally becomes
\begin{align}
    &\left[\mathrm{i}\partial_t \pm \lmk p + \frac{f}{2} \rmk\right] \psi^{(+,+)}_{\rm L/R}(t, {\bm p}) = 0, \label{su2DiracEq1}\\
    &\left[\mathrm{i}\partial_t \pm \lmk -p + \frac{f}{2} \rmk\right] \psi^{(-,-)}_{\rm L/R}(t, {\bm p}) = 0,\label{su2DiracEq2}\\
    &\left[\mathrm{i}\partial_t \pm \left(p\left( 
\begin{array}{cc}
1 & 0\\
0 & -1\\
\end{array}
\right) + \frac{f}{2} \lmk 
\begin{array}{cc}
-1 & 2\\
2 & -1\\
\end{array}
\rmk\right)
\right] 
\lmk
\begin{array}{c}
\psi^{(+,-)}_{\rm L/R}(t, {\bm p})\\
\psi^{(-,+)}_{\rm L/R}(t, {\bm p}) 
\end{array}
\rmk
= 0, \label{su2DiracEq3}
\end{align}
where, once more, $+ (-)$ refers left- (right-)handed fermions, respectively.
Since $(+,-)$ and $(-,+)$ modes are mixed, one needs to perform diagonalization to identify the energy eigenstates. After diagonalization, one finds the expression of energy dispersion relation for four independent modes as
\begin{equation}
\begin{aligned}
    \omega^{(+1)}_{\rm L/R} &= \pm\lmk -p - \frac{f}{2}\rmk,\\
    \omega^{(-1)}_{\rm L/R} &= \pm\lmk p - \frac{f}{2}\rmk,\\
    \omega^{(0;1)}_{\rm L/R} &= \pm \lmk -\sqrt{p^2 + f^2} + \frac{f}{2}\rmk,\\
    \omega^{(0;2)}_{\rm L/R} &= \pm\lmk \sqrt{p^2 + f^2} + \frac{f}{2}\rmk, 
\end{aligned}\label{eq:su2_dispersion}
\end{equation}
where  $+ (-)$ refers left- (right-)handed fermions, respectively.
Among these four modes, three modes are gapped and only $\omega^{(-1)}_{\rm L/R}$ smoothly connects the positive frequency state and negative frequency state at $p=f/2$.

\begin{figure}[htbp]
\begin{center}
\includegraphics[width=0.5\columnwidth]{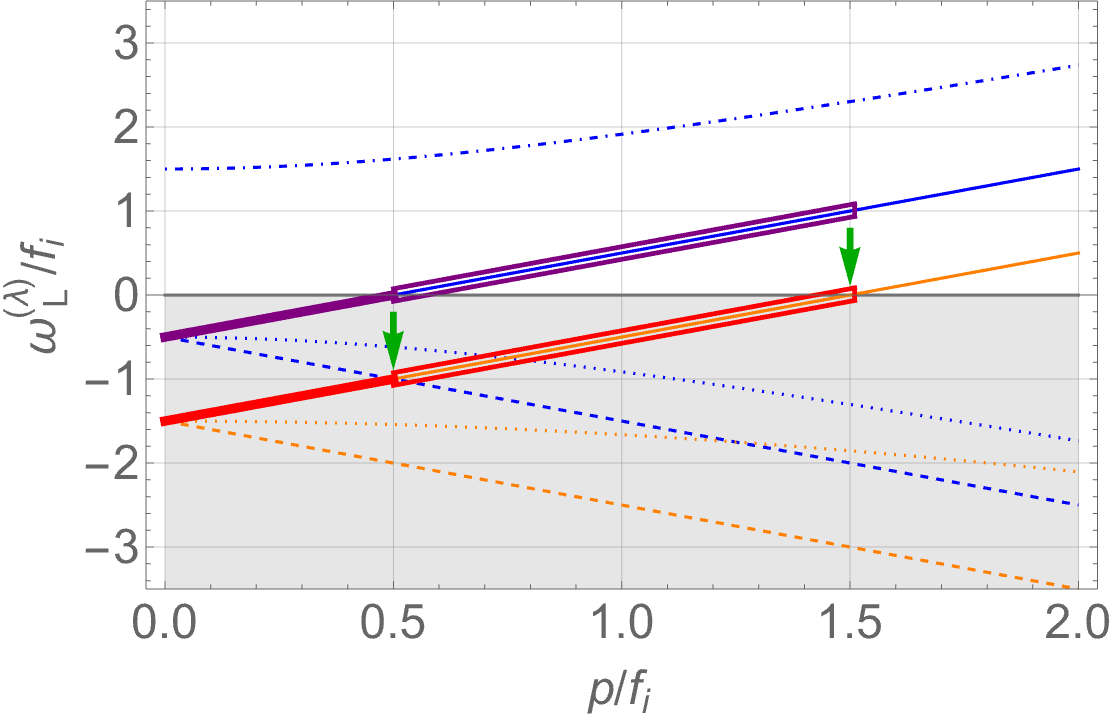}~
\includegraphics[width=0.5\columnwidth]{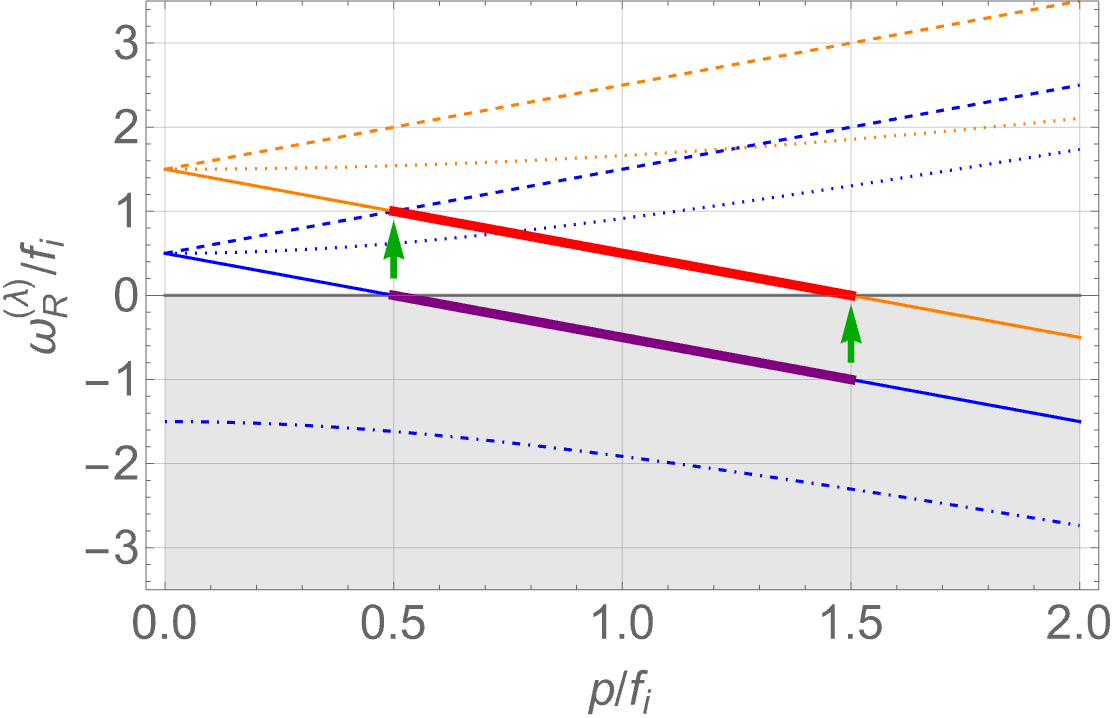}
\caption{\small
The levels for the massless left- (left figure) and right-handed (right figure) fermions in the background
field configuration of SU(2) gauge field~\eqref{configCN}, respectively. 
Blue lines are those for $f=f_i$ while 
orange lines are those for $f=f_f = 3 f_f$.
Solid, dashed, dotted, and dot-dashed lines represent 
$\omega^{(-1)}_{\rm L/R}, \omega^{(+1)}_{\rm L/R}, \omega^{(0;1)}_{\rm L/R}$, and $\omega^{(0;2)}_{\rm L/R} $, respectively. 
Gray shaded region represents the Dirac sea. 
While for the right-handed fermions states of $\omega=\omega^{(-1)}_{\rm R}$ from $p=f_i/2$ to $p=f_f/2$ in the Dirac sea at $f=f_i$ (red thick line) are excited at $f=f_f$
(purple thick line), 
for the left-handed fermions unoccupied states of $\omega=\omega^{(-1)}_{\rm }$ from $p=f_i/2$ to $p=f_f/2$ at $f=f_i$ (red double line) become holes at $f=f_f$
(purple double line). 
Note that the horizontal axes are those for the physical momenta 
but not the conjugate momenta, 
and hence we do not see the sideways shift of the momentum 
seen in the U(1) case. }
\label{fig:SU2LC}
\end{center}
\end{figure}

By adiabatically evolving $f$ from $f_i$ (at $t=t_i$)  to $f_f$ (at $t=t_f$)  , level crossing occurs for these lowest modes as in the Abelian case where an electric field is applied to LLL.
Figure~\ref{fig:SU2LC} shows the schematic picture of the level crossing in this system.
With the normally ordered charge operator, the chiral charge associated with this excitation can be evaluated as
\begin{align}
    \Delta Q_{5}^{(\mathrm{e})} =\Delta Q_{\rm R}^{(\mathrm{e})} - \Delta Q_{\rm L}^{(\mathrm{e})}  =  
    \frac{1}{3} \times V\frac{1}{8\pi^2}\lmk f_f^3 - f_i^3 \rmk,
\end{align}
where $V \equiv \int \mathrm{d}^3x$ is the volume factor, and 
we have taken into account that in the momentum space there is one state occupied in each volume $V \mathrm{d}^3p/(2\pi)^3=1$. 
Since we take the infinite volume limit, we do not discretize the momentum space.
In contrast to U(1) case, however, this contribution from excitation does not account for all the fermion chirality:
\begin{align}
    \Delta\langle Q_{5}\rangle = \Delta (\langle Q_{\rm R} \rangle -\langle Q_{\rm L} \rangle ) = V\frac{1}{4\pi^2}\lmk f_f^3 - f_i^3 \rmk, \label{eq:su2_net}
\end{align}
which is derived by integrating anomaly equation, 
\begin{equation}
    \partial_\mu J^\mu_5 = -\frac{1}{16 \pi^2} F_{\mu\nu}^a \tilde{F}^{a\mu\nu} =  \frac{1}{4 \pi^2} (f^3)^\cdot,
\end{equation}
from $t_i$ to $t_f$ and over whole space. 
This is because one needs to take into account the vacuum contribution of chiral charge, known as the eta invariant in the Atiyah-Patodi-Singer index theorem~\cite{Atiyah:1963zz,Atiyah:1968mp}. The theorem indicates the following relation of the charges in our Lorentzian manifold~\cite{Gibbons:1979kq,Stone:2023qln}:
\begin{align}
    \Delta\langle Q_{5}\rangle = \int \mathrm{d}t \mathrm{d}^3 x \partial_\mu  J^\mu_5 &= \Delta Q_{\rm R}^{(\mathrm{e})} - \Delta Q_{\rm L}^{(\mathrm{e})} - [\eta_H]^{t=t_f}_{t=t_i},\\
    &= \Delta Q_{\rm R}^{(\mathrm{e})} - \Delta Q_{\rm L}^{(\mathrm{e})} + \Delta Q_{\rm R}^{(\mathrm{v})} - \Delta Q_{\rm L}^{(\mathrm{v})},
\end{align}
where the vacuum charge can be identified as $\Delta Q_{\rm R/L}^{(\mathrm{v})} = \mp (1/2)[\eta_H]^{t=t_f}_{t=t_i}$.
Here $\eta_H$ is the eta invariant defined with respect to the Hamiltonian of (right-handed) Weyl fermion as
\begin{equation}
    \eta_H = \lim_{s \rightarrow 0} \eta_H(s), \quad \eta_H(s) \equiv \sum_{\omega_{\rm R} \not =0} \mathrm{sgn}(\omega_{\rm R}) |\omega_{\rm R}|^{-s} = -\sum_{\omega_{\rm L} \not =0} \mathrm{sgn}(\omega_{\rm L}) |\omega_{\rm L}|^{-s}, 
\end{equation}
which diverges and requires an appropriate regularization.
In Ref.~\cite{Domcke:2018gfr}, it is shown that
the chiral anomaly equation is reproduced with the regularized vacuum contribution
\begin{align}
    \Delta Q_{\rm R/L}^{(\mathrm{v})} = \lkk\lim_{\Lambda \rightarrow \infty} -\frac{1}{2}
 \sum_{\omega_{\rm R/L} \not = 0} {\rm sgn} (\omega_{\rm R/L}) R\left( \frac{\omega_{\rm R/L}}{\Lambda}\right)\rkk^{t=t_f}_{t=t_i}
\end{align}
where the regulator $R(s)$ is smooth and approaches rapidly to zero sufficiently, and satisfies $R(s\to0) = 1$.
By substituting the dispersion~\eqref{eq:su2_dispersion}, chiral charge associated with vacuum contribution is evaluated as
\begin{align}
    \Delta Q_{5}^{(\mathrm{v})} &= \Delta Q_{\rm R}^{(\mathrm{v})} - \Delta Q_{\rm L}^{(\mathrm{v})}\\
    &=[\eta_H]^{t=t_f}_{t=t_i} = \frac{5}{6}  \times V\frac{1}{4\pi^2}\lmk f_f^3 - f_i^3 \rmk,
\end{align}
with which the net chirality $\Delta\langle Q_{5}\rangle = \Delta Q_{5}^{(\mathrm{e})} + \Delta Q_{5}^{(\mathrm{v})}$ becomes consistent with the prediction of anomaly equation.
Let us stress that in contrast to the case of U(1) gauge field discussed above, the contribution from gapped modes $\omega^{(+1)}_{\rm L/R}, \omega^{(0;1)}_{\rm L/R}$, and $\omega^{(0;2)}_{\rm L/R} $ participates in $\Delta\langle Q_{5}\rangle$ via $\Delta Q_{5}^{(\mathrm{v})}$.

Physically, the eta invariant accounts for the chiral asymmetry accumulated in the vacuum due to the asymmetric energy spectrum of fermions. 
It is quite interesting that a substantial fraction of the chiral charge lies in the vacuum in this SU(2) gauge field configuration.
At this point, one may wonder whether the charge accumulated in the vacuum plays a physical role or not.
In other words, it is not clear how this charge interacts with other fields or particles, if ever.
For example, there also exists a vacuum contribution in the SU(2) current (not the chiral current) of the fermion in this system. The authors of Ref.~\cite{Domcke:2018gfr} has shown that such a contribution can be renormalized with the running coupling constant and does not interact with other fields.
As a result, the backreaction from the fermion to the SU(2) gauge fields is turned out to be inefficient in contrast to the case of U(1) gauge field.
Anyway, the above observation indicates that if one needs to distinguish the excitation and vacuum contribution, one should not naively use the integrated anomaly equation~\eqref{eq:su2_net}.
As we will see, this distinction could be more important for the gravitational anomaly
and leptogenesis.

\subsection{Lessons from gauge field cases}\label{sec:lesson}
Before proceeding, we summarize what we learn from the 
chirality production in terms of the level crossings
in the cases of the U(1) and SU(2) gauge fields. For both cases, degeneracy in the helicity and the spin states are broken by the presence of homogeneous magnetic field (or constant $f$ with the non-trivial configuration in the SU(2) case). 
Consequently, the ``lowest'' mode globally appears, where the spin polarization is along with the magnetic component of the gauge field, smoothly connecting the negative and positive frequency modes.
The generation of chirality is, however, qualitatively different.
For the U(1) case, only this lowest mode participates in the chirality generation. 
That is, the electric field causes momentum shift in LLL and the excited states account for all the chirality generated in this system. 
On the contrary, the (higher) gapped modes also contribute in the homogeneous and isotropic SU(2) case as the vacuum contribution although level-crossings never take place there. In this case, the amplitude of the SU(2) field value $f$ plays a similar role as the chiral chemical potential and causes chirality dependent bias for all the modes. Consequently, the evolution of bias $f$ (or the ``electric component'' of SU(2) field) results in the accumulation of vacuum chiral charge, including the modes without level-crossing. This is the most striking difference from the U(1) example.

Although it is sub-leading, the excited states still have non-negligible contribution for SU(2) case. In this respect, we conjecture that for having non-negligible contribution from excitation, homogeneity of the external field seems to play a key role. Because of the homogeneity, the modes relevant for the chiral particle production are globally defined over the whole momentum space, both for the U(1) and SU(2).
This global nature allows fermions to have continuous excitation as long as the electric field or growth of $f$ is provided.
As we will see in the following, this seems the crucial difference from the ``inhomogeneous and anisotropic'' chiral spin-2 gravitational field we consider in Sec.~\ref{sec:weak_spin2}.
As a supporting example of this discussion, let us refer to Ref.~\cite{Christ:1979zm} where the author evaluated the fermion chirality production under the gauge field ``radiations'' carrying topological charge. Interestingly, for the U(1) case, the vacuum charge accounts for all the chirality generated in this system.
Moreover, the U(1) chiral anomaly is recently investigated under the inhomogeneous electric field, which also shows that chiral charge is not produced\footnote{In this case, anomaly equation is saturated by the generation of $J_5^1$. Although this is qualitatively different from our case where $\langle J_5^0 \rangle \neq 0$ dominated by vacuum contribution, it still suggests the difficulty of exciting $J_5^0$ in a non-uniform external field.}~\cite{Fukushima:2023obj}.

Another thing we would like to mention here is that, as discussed in Ref.~\cite{Domcke:2018gfr}, there are two conditions for vacuum contribution to be non-vanishing. 
I) The initial and final external field configuration must not be equivalent up to gauge transformation. 
II) By definition, the spectrum of positive and negative frequency modes must be asymmetric. 
Both of them are not satisfied in the case of homogeneous U(1) electromagnetic field with infinitely long magnetic flux.
In order to satisfy the first condition, evolution in the magnetic component of the field is indispensable.
If these conditions are satisfied while the excitation does not reproduce the anomaly equation, we can presume that the vacuum contribution would compensate to reproduce the anomaly equation.
Note that {\it e.g.} when a system with a finite loop of the magnetic flux is concerned, $\eta$ invariant becomes essential in fulfilling the anomaly even for the U(1) gauge field background (see App.~\ref{sec:u1_s1}).
Such a non-trivial manifold is, however, not under consideration here, since we are interested in particle production in the spatially flat expanding universe.

As we will see below, the first condition for the existence of the vacuum contribution is always satisfied in the chiral GW generation. In this situation, one may naturally come up with the following question: Can the vacuum contribution be also dominant in gravitational leptogenesis?
In fact, the dominance of vacuum asymmetry is observed in the gravitational system, for example, Bianchi type-IX~\cite{Gibbons:1979kq, Gibbons:1979ks} and Bianchi type-II~\cite{Stone:2023qln} spacetimes.
In the following section, before investigating the possibility of level crossing in the gravitational leptogenesis, we discuss the fermion chirality production in the Bianchi type-IX spacetime to see the similarity to the above example of SU(2) gauge field.

\section{A solvable exmaple in gravitational system: Bianchi type-IX spacetime}\label{sec:bianchi}
Now let us turn to the chirality production 
under the non-trivial gravitational background. 
Before examining the configuration motivated from the gravitational leptogenesis, we first review the case with closed homogeneous spacetime, classified according to the Bianchi type models.

Among the Bianchi classification of the homogeneous $3 + 1$ spacetime, Bianchi type-IX is particularly of our interest.
As discussed in Refs.~\cite{Grishchuk:1974ny, King:1991jd}, this spacetime can be expressed as a closed FLRW spacetime onto which circularly polarized GWs are superimposed.
This fact motivates us to investigate the gravitational chiral anomaly in this system since the circular polarization of GWs yields non-vanishing Chern-Pontryagin density. 
Indeed, Gibbons has discussed gravitational production of neutrino in the evolving spacetime which starts from ``polarized'' initial state to the isotropic (unpolarized) final state~\cite{Gibbons:1979ks, Gibbons:1979kq}.
Let us give a review on how the chirality of fermion is generated in this system with certain clarification compared to the previous studies.

\subsection{Bianchi type-IX and GW interpretation}
We first briefly review the general characteristics of Bianchi type-IX spacetime, following the discussion in Ref.~\cite{King:1991jd}. 
A right-homogeneous Bianchi type-IX spacetime is generally described by a metric (with the Cartan calculus) as 
\begin{align}
    \mathrm{d}s^2 & = -a^2(\eta)  (\mathrm{d} \eta^2 - \mathrm{d}l^2),\label{genmetricCC}   \\
    \mathrm{d}l^2 & = A_{ij} (\eta) \lambda^i \otimes \lambda^j, \label{genspacemetricCC}
\end{align}
where $\eta$ is the conformal time, and $\lnk\lambda^i\rnk$ are three left-invariant 1-form on the 3-sphere $S^3$,\footnote{On $S^3$, there are two different ways to construct three translational Killing vectors $\lnk\xi_{(1)},\xi_{(2)},\xi_{(3)}\rnk$. Depending on the sign of their commutation relation $\lkk \xi_{(i)}, \xi_{(j)}\rkk \propto \mp \epsilon_{ijk}\xi_{(k)}$, they are referred to as the left-translation (-) and right-translation (+), respectively. 1-form (or the measure of the 1-dim integration) invariant under this left-translation is called as left-invariant. 
For the left-invariant 1-forms~\eqref{dleft1}-\eqref{dleft3}, the corresponding killing vectors satisfy the commutation relation $\lkk \xi_{(i)}, \xi_{(j)}\rkk = - (2/S) \epsilon_{ijk}\xi_{(k)}$.~\label{footnotesu2}} satisfying
\begin{equation}
    \mathrm{d} \lambda^i = \frac{1}{S}\epsilon_{ijk} \lambda^j \wedge \lambda^k. \label{ExDer1f}
\end{equation}
More concretely, they can be expressed as
\begin{align}
\lambda^1& =  \frac{S}{2} \left[ - \sin \psi \mathrm{d} \theta + \sin \theta \cos \psi \mathrm{d} \phi \right], \label{dleft1}\\
\lambda^2 & =\frac{S}{2} \left[ \cos \psi  \mathrm{d} \theta + \sin \theta \sin \psi \mathrm{d} \phi \right], \label{dleft2}\\
\lambda^3 & = \frac{S}{2} \left[ \cos \theta \mathrm{d} \phi + \mathrm{d} \psi \right],   \label{dleft3}
\end{align}
with $(\theta, \psi, \phi)$ being the Euler angles which run $0 \leq \theta \leq \pi, \ 0 \leq \psi \leq 2\pi, \ 0 \leq \phi \leq 4\pi$~\cite{Misner:1969hg}, and $S$ being the (comoving) radius of $S^3$. 
The spatial part of the metric~\eqref{genspacemetricCC} can be decomposed into the background closed sphere (or the closed FLRW Universe) and the five independent tensor fields as 
\begin{equation}
A_{ij} (\eta) = A_0(\eta)\delta_{ij} + \sum_{a = 1}^5A_a(\eta)C^{a}_{ij},\label{eq:type-ix_basis}
\end{equation}
where $C^{a}_{ij}$ are five linearly independent traceless matrices. 
While the isotropy of background sphere is broken, the tensor fields $A_a(\eta)C^{a}_{ij}\lambda^i \otimes \lambda^j$ imposed onto the sphere preserves its right homogeneity.
In this sense, one can understand that the Bianchi type-IX spacetime is an anisotropic generalization of the closed FLRW Universe.

Another aspect of this spacetime we would like to mention is that this anisotropy of closed sphere involves the parity-violation.
As shown in Ref.~\cite{King:1991jd}, the right homogeneous tensor fields described by $A_a(\eta)C^{a}_{ij}\lambda^i \otimes \lambda^j$ are the 
symmetric, transverse, and traceless tensor spherical harmonics on the background 3-sphere.
In other words, they can be understood as the standing GWs around the sphere with the longest wavelength. 
In addition, these waves have the left-circular polarization (see Ref.~\cite{King:1991jd} for graphical description).
Therefore, the right homogeneous Bianchi type-IX spacetime can be decomposed into the background closed sphere plus left circularly polarized GWs~\cite{Grishchuk:1974ny,King:1991jd}. Note that the same discussion applies to the left homogeneous Bianchi type-IX and in this case the spacetime is wrapped by the right circularly polarized GWs.
As we will see below with a specific example of the metric, parity-violating GWs yield non-vanishing Chern-Pontryagin density. Therefore, this spacetime can be a playground to investigate the generation of fermion chirality.

Now we consider a specific class of this spacetime called as the axial Bianchi type-IX, which was considered in Refs.~\cite{Gibbons:1979kq,Gibbons:1979ks} to examine the chiral fermion production. 
The metric of the axial Bianchi type-IX is given as 
\begin{align}
    \mathrm{d}s^2 
    &=-a(\eta)^2 \left(\mathrm{d}\eta^2 - (e^{2\beta_+(\eta)}\lambda^1\otimes\lambda^1 + e^{2\beta_+(\eta)}\lambda^2\otimes\lambda^2 + e^{-4\beta_+(\eta)}\lambda^3\otimes\lambda^3) \right), \notag \\
    &= - a(\eta)^2 \left(\mathrm{d}\eta^2 - \frac{S^2}{4}\left[e^{2\beta_+(\eta)}\mathrm{d}\theta^2 + \lmk e^{2\beta_+(\eta)}\sin^2\theta + e^{-4\beta_+(\eta)}\cos^2\theta\rmk \mathrm{d}\phi^2 \right. \right.\notag \\
    & \left.\left.  \ \ \ \ \  + 2e^{-4\beta_+(\eta)}\cos \theta \mathrm{d}\phi \mathrm{d}\psi + e^{-4\beta_+(\eta)}\mathrm{d}\psi^2 \right] \right),\label{bianchi_inv}
\end{align}
where $\beta_+$ is a function of time and characterizes the anisotropy of the closed sphere. At the same time, $\beta_+$ also characterizes the amplitude of the polarized GWs. This can be seen by performing the decomposition mentioned above: 
\begin{align}
    \mathrm{d}s^2 = -a(\eta)^2 \left(\mathrm{d}\eta^2 -   \left(\mathrm{d}l^2_{B} + \mathrm{d}l^2_h \right) \right),
\end{align}
where $\mathrm{d}l^2_{B}$ is the background sphere, 
\begin{align}
    \mathrm{d}l^2_{B} &= \frac{(2e^{2\beta_+} + e^{-4\beta_+})}{3} \delta_{ij} \lambda^i \otimes \lambda^j \notag \\
    &=\frac{(2e^{2\beta_+} + e^{-4\beta_+})}{3}\frac{S^2}{4}\left[\mathrm{d}\theta^2 + \mathrm{d}\phi^2+ 2\cos \theta \mathrm{d}\phi \mathrm{d}\psi + \mathrm{d}\psi^2 \right].
\end{align}
On the other hand, $\mathrm{d}l^2_h$ is the standing GW described by one specific mode ($a = 5$ in Eq.~\eqref{eq:type-ix_basis} if we follow the notation in Ref.~\cite{King:1991jd})
\begin{equation}
    \frac{1}{\sqrt{6}} \left(- \lambda^1 \otimes \lambda^1 -\lambda^2 \otimes \lambda^2 + 2\lambda^3 \otimes\lambda^3 \right)\label{q35}
\end{equation}
as
\begin{align}
 \mathrm{d}l^2_h &= \frac{(e^{2\beta_+} - e^{-4\beta_+})}{3} \left( \lambda^1 \otimes \lambda^1 + \lambda^2 \otimes \lambda^2 - 2\lambda^3 \otimes\lambda^3\right) \notag \\ 
 &=\frac{(e^{2\beta_+} - e^{-4\beta_+})}{3}\frac{S^2}{4}\left[\mathrm{d}\theta^2 + (1 - 3 \cos^2 \theta) \mathrm{d} \phi^2 -4 \cos \theta \mathrm{d}\phi \mathrm{d}\psi -2 \mathrm{d} \psi^2 \right] . 
\end{align} 
Note that the standing GW is not restricted to be perturbative as ordinary GWs. Hence, this expression in the large anisotropy limit $\beta_+ \gg 1$ still makes sense.

For this metric~\eqref{bianchi_inv}, the gravitational Chern-Pontryagin density $R\tilde{R}$ is evaluated as
\begin{align}
    R\tilde{R} &= -\frac{48}{a^4 S^3} \beta_+^{\prime}\lmk 4k^4 - 4k^2 -  6 S^2 k^{4/3}\beta_+^{\prime 2}+ 3S^2 k^{4/3}\beta_+^{\prime\prime}\rmk \notag \\
    &=\frac{1}{\sqrt{-g}} \partial_\eta \left( 6 \sin \theta \left(\frac{1}{3}(k^2 - 1)^2 -\frac{3S^2}{2}k^{4/3}\beta_+^{\prime 2}  \right)\right)  , 
    \quad \text{with} \quad k \equiv \exp(-3\beta_+), \label{gravCAE}
\end{align}
where $g = - S^6 a^8 \sin^2 \theta/64$ is the determinant of the metric and the prime denotes the derivative with respect to the conformal time. 
Here we have introduced a new parameter $k$ following Refs.~\cite{Gibbons:1979kq,Gibbons:1979ks}. 
One can see that $R\tilde{R}$ becomes non-vanishing when the anisotropy $\beta_+$ (or $k$) evolves.
This is a clear indication of parity-violation due to the polarized GW, which may be quantified by the gravitational helicity density defined as 
\begin{equation}
   \sqrt{-g} K^0 = 6 \sin \theta \left(\frac{1}{3}(k^2 - 1)^2  -\frac{3S^2}{2}k^{4/3}\beta_+^{\prime 2}  \right).
\end{equation}
Since the spacetime with different $\beta_+$ or $k$ is 
not equivalent up to the gauge transformation, 
the first condition for the existence of the vacuum contribution in the particle production is satisfied.
This motivates us to investigate the particle production
in terms of the level crossing to clarify the distribution of chiral charge. 

\subsection{Chirality production and comparison to the SU(2) gauge example}
We now see the chirality production in the point
of view of the level crossing.
According to the gravitational chiral anomaly equation, 
\begin{equation}
   \nabla_\mu J^\mu_5 = \frac{1}{\sqrt{-g}} \partial_\mu (\sqrt{-g} J^\mu_5) = -\frac{1}{12 (4 \pi)^2}R\tilde{R},
\end{equation}
we would expect that the fermion chirality production  takes place when $\beta_+$ (or $k$) changes.
Integration of the chiral anomaly equation~\eqref{gravCAE} leads to
\begin{align}
\Delta Q_5 & = - \int_0^\pi \mathrm{d}\theta \int_0^{2 \pi} \mathrm{d}\psi \int_0^{4 \pi} \mathrm{d}\phi   \frac{\sin \theta }{2(4\pi)^2}  \Delta \left(\frac{1}{3}(k^2 - 1)^2 -\frac{3S^2}{2}k^{4/3}\beta_+^{\prime 2}  \right) \notag \\
&=-\frac{1}{6} \Delta \lmk (k^2 - 1)^2 -3 k^{4/3} (\beta'[\eta])^2 \rmk.\label{eq:anomaly_integ_bianchi}
\end{align}
Then we wonder how they
are divided into the excitation and vacuum contribution.
Following Refs.~\cite{Gibbons:1979kq,Gibbons:1979ks}, let us discuss the generation of fermion chirality under the adiabatic evolution of the spacetime assuming $\beta_+^{\prime} \to 0$.

Since the Bianchi-IX space has an SU(2) structure as seen in Eq.~\eqref{ExDer1f},
it is convenient to study the system 
in the frame with the Cartan calculus (see Eq.~\eqref{genspacemetricCC}). 
The Dirac equation for the massless fermion in that frame is given as~\cite{Brill:1966tia}
\begin{equation}
\mathrm{i} \gamma^\mu (\omega_\mu(\Psi) - \Gamma_\mu \Psi) =0, \label{DiracEqBrill}
\end{equation}
where $\omega_\mu$ are the dual of the 
normalized 1-forms in 4 dimension $\omega^\mu$, such that it forms Minkowskian metric. The connection $\Gamma_\mu$ is defined as 
\begin{equation}
    \Gamma_\mu = -\frac{1}{4} \omega_{\mu\nu\rho} \gamma^\nu \gamma^\rho,
\end{equation}
where $\omega^\mu_{\ \ \nu\rho}$ satisfies $d \omega^\mu = \omega^\mu_{\ \ \nu \rho} \omega^\nu \wedge \omega^\rho$.

\begin{figure}[htbp]
\begin{center}
\includegraphics[width=0.6\columnwidth]{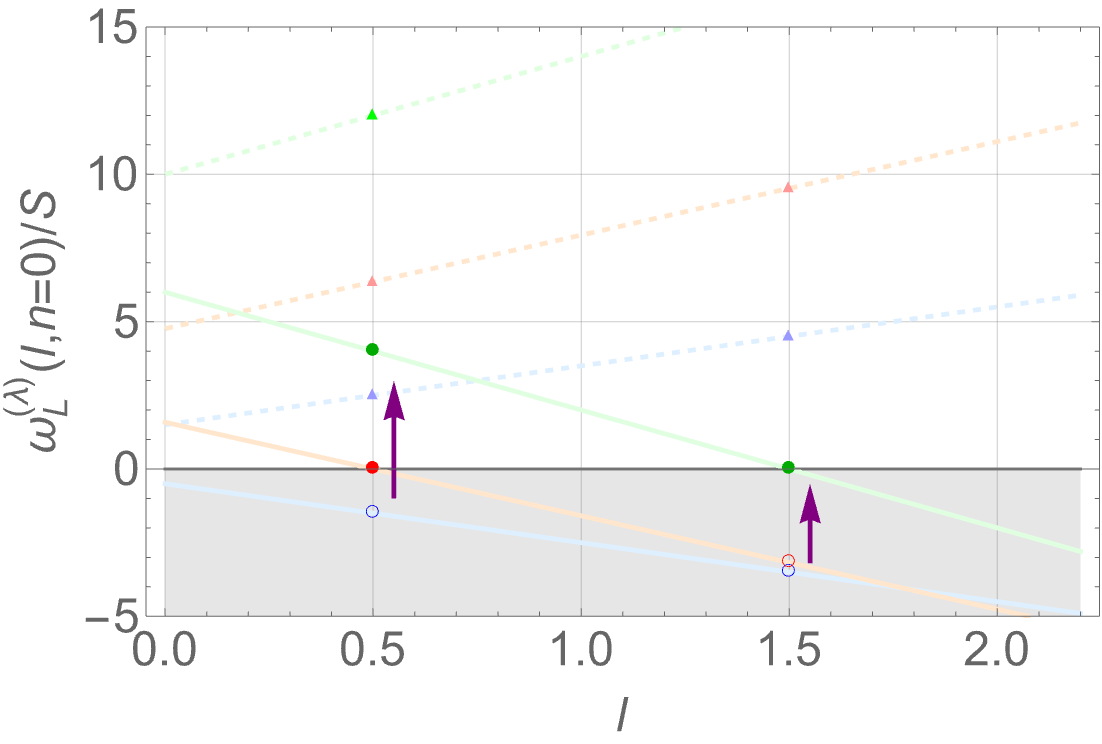}
\caption{\small The levels for the massless left-handed fermions in the Bianchi IX 
spacetime~\eqref{bianchi_inv} for $n=0$. Blue, red, and green lines and markers are those for
$k=1,4, 8$, respectively. Solid lines and circles represent $\omega_{\rm L}^-$ while dashed lines and triangles represent
$\omega_{\rm L}^+$. Gray shaded region represents the Dirac sea. 
While all the states of  $\omega_{\rm L}^-$ are in the Dirac sea for $k=1$, 
the highest state $l=1/2$ appears as the excitation at $k=4$
while the state $l=3/2$ appears at $k=8$. 
 }
\label{fig:BianchiIXLC}
\end{center}
\end{figure}

In our case of Eq.~\eqref{bianchi_inv}, the Dirac equation can then be expressed as
\begin{align}
(\mathrm{i} \partial_{\eta} \mp D)(a^{3/2}(\eta)\psi_{L/R}) = 0,\label{Dirac_BIX}
\end{align}
where - (+) refers left- and right-handed fermions, respectively, and 
the Dirac operator $D$ is defined as\footnote{The expression of the Dirac operator is different from the one in Ref.~\cite{Gibbons:1979kq,Gibbons:1979ks} due to the difference in the choice of the coordinate basis, but the physics is unchanged. }
\begin{align}
D &= \frac{k^{1/3}}{S}  \left( \begin{array}{cc} \dfrac{2 L_{(3)}}{k} +  \left(\dfrac{1}{k}+\dfrac{k}{2}  \right) & 2 L_- \\  2 L_+ & -\dfrac{2 L_{(3)}}{k} + \left(\dfrac{1}{k}+\dfrac{k}{2}  \right)\end{array}\right). \label{Dirac_op}
\end{align}
Here we have dropped the terms proportional to $\beta_+^{\prime}$ and defined 
\begin{equation}
L_{(i)}  \equiv - \frac{\mathrm{i}S}{2} \xi_{(i)}, \quad L_\pm = L_{(1)}\pm \mathrm{i} L_{(2)}. 
\end{equation}
The constant terms  $(1/k+k/2)$ comes from the connection,
while terms with $L_{(i)}$ comes from $\omega_\mu$. 
Note that $\{L_{(i)}\}$ satisfy the spin algebra, 
\begin{equation}
    [L_{(i)}, L_{(j)}] = \mathrm{i} \epsilon_{ijk} L_{(k)}, \quad [L_{(3)},L_\pm] = \pm L_\pm, 
\end{equation}
and hence $L_\pm$ can be understood as the ladder operators. 
Then we can introduce eigenfunction $\left|l, m , n\right \rangle$ as 
\begin{align}
    L_3\left|l, m , n\right\rangle &= n\left|l, m , n\right\rangle,\\
    L_{\pm}\left|l, m , n\right\rangle &= \sqrt{(l\mp n)(l \pm n + 1)}\left|l, m , n\pm1\right\rangle,
\end{align}
where $l$ is a non-negative half-integer or a natural number, and $m,n$ are (half-) integers satisfying $m, n = -l, -l + 1, ..., l - 1, l$.
Here the quantum number $m$ is introduced because 
these states are also representations of the right-translations.
Using this function, eigenstates of the Dirac operator~\eqref{Dirac_op} can be constructed as
\begin{align}
    \Phi_{\pm} (l,m,n) = 
\left(\begin{array}{c} \left(n/k \pm \sqrt{(n/k)^2 + (l+1/2)^2-n^2} \right)|l,m,n-1/2 \rangle \\ 
 \sqrt{(l+1/2)^2-n^2}  |l,m,n+1/2 \rangle 
\end{array}\right), \label{eigen_func}
\end{align}
whose eigenvalues are
\begin{equation}
    \frac{k^{1/3}}{S} \left(\dfrac{k}{2} \pm 2 \sqrt{\left(\dfrac{n}{k}\right)^2 + \left(l+\dfrac{1}{2}\right)^2 -n^2} \right),
\end{equation}
Here $m$ runs from $-l$ to $l$, which gives a degeneracy $2l+1$, while $n$ runs from $-l-1/2$ to $l+1/2$. 
Note that for $|n|=l+1/2$, we only have $\Phi_+(l,m,l+1/2)$ since $\Phi_- (l,m,l+1/2)$ is a null state. 
Thus we obtain the energy dispersion relation as
\begin{align}
    \omega_{\rm L}^\pm (l,m,n)&=  \frac{k^{1/3}}{S} \left(\dfrac{k}{2} \pm 2 \sqrt{\left(\dfrac{n}{k}\right)^2 + \left(l+\dfrac{1}{2}\right)^2 -n^2} \right), \label{eq:eigen_valuesL} \\
    \omega_{\rm R}^\pm (l,m,n)&= -  \frac{k^{1/3}}{S} \left(\dfrac{k}{2} \pm 2 \sqrt{\left(\dfrac{n}{k}\right)^2 + \left(l+\dfrac{1}{2}\right)^2 -n^2} \right). \label{eq:eigen_valuesR} 
\end{align}
for the left- and right-handed fermions, respectively. 
While $\omega_{\rm L}^+$ and $\omega_{\rm R}^+$ are positive and 
negative definite, respectively, $\omega_{\rm L}^-$ and $\omega_{\rm R}^-$ connects the positive and negative energy eigenstates. 
Figure~\ref{fig:BianchiIXLC} shows the dispersion relation 
of the left-handed fermions for $n=0$ and some choices of $k$. 

As can be seen from Eqs.~\eqref{eq:eigen_valuesL} and \eqref{eq:eigen_valuesR}, in the isotropic background ($k=1$), all the states of $\omega^-_{\rm L}$ are in the Dirac sea and those of $\omega^-_{\rm R}$ have positive definite energy eigenvalue.
The sign of eigenvalue $\omega^-_{\rm L/R}$ can be flipped for large enough $k$, or in other words, ``non-perturbatively'' large amplitude of the standing wave. 
For example, the lowest energy state $(l, n) = (1/2, 0)$ changes the sign at $k = 4$.
Therefore, if the universe evolving from the vacuum isotropic space to large $k$, left-handed fermions appear as excitations while right-handed fermions develop holes as a consequence of the zero-crossing, 
which leads to a negative chirality production (or vice versa as discussed in Refs.~\cite{Gibbons:1979ks, Gibbons:1979kq})\footnote{These are not solutions to the Einstein equation. Here the evolving spacetime is considered as an artificial external field to the fermions.}, 
This excited contribution of chiral charge is then evaluated as
\begin{equation}
  \Delta Q_5^{(\mathrm{e})} = \Delta Q_{\rm R}^{(\mathrm{e})} -\Delta Q_{\rm L}^{(\mathrm{e})} = - \sum_{l,n} 2 (2l+1),  
\end{equation}
where $(l,n)$ runs the states that experiences the level crossing. 
Here $2l+1$ counts the degeneracy of $m$ and 2 counts the left- and right-handed
fermions.

In the large $k$ limit, we can obtain an approximate evaluation by replacing the summation to the integral, $\int dl dn$, with the region of $(l,n)$,
\begin{align}
    |n| \leq l - \frac{1}{2},     \quad\lmk l + \frac{1}{2}\rmk^2 - n^2 < \frac{k^2}{16}, 
\end{align}
which leads to
\begin{equation}
    \Delta Q_5^{(\mathrm{e})}   \simeq - \frac{k^4}{256}. 
\end{equation}
The $k$ dependence, $\Delta Q_5^{(\mathrm{e})}  \propto k^4$, is the same to the 
anomaly equation~\eqref{eq:anomaly_integ_bianchi}. 
This means that the level crossing contribution continues to be non-negligible with respect to $\Delta Q_5$ even in $k\to\infty$ limit. 
We anticipate that this feature is due to the homogeneity of system as we discussed in Sec.~\ref{sec:lesson}. That is, the system admits globally defined momentum (or the infinite sets of label $(l, m, n)$) over phase space and also global selection of spin states.
Therefore, as long as the anisotropy increases (or decreases), there will always be modes excited as in the gauge field examples.

There is, however, an order of magnitude difference between $\Delta Q_5^{(\mathrm{e})}$ and $\Delta Q_5$ predicted from the anomaly equation~\eqref{eq:anomaly_integ_bianchi}. 
Moreover at $k<4$ there are no level crossing at all as has been seen in Fig.~\ref{fig:BianchiIXLC}, while the anomaly equation~\eqref{eq:anomaly_integ_bianchi} suggests the generation of the chiral asymmetry for any $k>1$. This indicates that the chirality is not mainly carried by the particles (excitation) but the vacuum.
Indeed, Ref.~\cite{Hitchin:1974rbi} has shown that for $k<4$ the eta invariant
is evaluated (see also Ref.~\cite{Stone:2023qln}) as
\begin{align}
   \eta_H(0) = -\frac{1}{6}(1 - k^2)^2,  \label{eq:eta_bianchi} 
\end{align}
while larger $k$ we shall subtract the (sub-dominant) contribution of the level-crossing.
Since the eta invariant in the small anisotropy regime~\eqref{eq:eta_bianchi} fully agrees with 
the chiral asymmetry predicted by the anomaly equation \eqref{eq:anomaly_integ_bianchi}, 
we safely conclude that the anomaly equation holds by taking into account
both the excitation and vacuum contribution, 
\begin{equation}
    \Delta Q_5 = \Delta Q_5^{(\mathrm{e})} + \Delta Q_5^{(\mathrm{v})}, \quad \Delta Q_5^{(\mathrm{v})}  = [\eta_H(0)]_{t = t_i}^{t = t_f}. 
\end{equation}

We now obtain further qualitative insights on the chiral anomaly in this system by comparing to the homogeneous SU(2) gauge field case. 
From Eqs.~\eqref{eq:eigen_valuesL} and \eqref{eq:eigen_valuesR}, one can easily see that the spatial anisotropy $k$ behaves similar to the chemical potential $\mu_{\rm R/L}$ and breaks the degeneracy as it causes energy bias $\sim k^{4/3}/2S$ between the positive and negative energy mode.
This is similar to the case with homogeneous SU(2) gauge field discussed in Sec.~\ref{sec:gauge_su2}, where the energy bias $f/2$ leads to chirality production in a very different way from the Abelian case.
It is interesting to note that the bias term in the Dirac operator~\eqref{Dirac_op} comes from the connection term in Eq.~\eqref{DiracEqBrill} and can be regarded as the ``non-Abelian'' contribution of the homogeneous but anisotropic geometry.
Note that the bias term exists even for the background sphere without GWs, $\beta_+=0$ or $k=1$.
Therefore, one might expect that such a chemical potential-like bias could arise if the ``non-Abelian'' nature is inherent to the spacetime geometry, which is already curved at the background.\footnote{One may seek the origin of such a similarity of Bianchi type-IX to the SU(2) gauge field example for the SU(2) algebra of momentum globally defined over the closed sphere. 
Indeed, in the both cases, the eigenstates (or the solution to the Dirac equation) are constructed similarly to the addition of angular momentum.
The similar bias is, however, also observed in the case of Bianchi type-II spacetime~\cite{Stone:2023qln}, where the momentum is no longer associated with the SU(2) algebra.}.
Such a concordance reminds us the sub-leading (but non-negligible) contribution from the smooth excitation and the domination of vacuum contribution in the SU(2) case.
Then, the distribution of chiraliy in the Bianchi type-IX discussed above is no longer surprising to us.

Let us summarize this section with a comment on the GW interpretation of the Bianchi type-IX spacetime.
If one assumes $|\beta_+| \ll 1$, the amplitude of gravitational standing waves becomes small enough and they seems to resemble the ``ordinary'' GWs which are the fluctuation around the ``flat'' spacetime. As we will discuss below, however, the ordinary GWs are essentially different from the standing GWs around closed sphere. 
Therefore, while the fact that level-crossing does not occur unless the amplitude of standing wave is non-perturbatively large is suggestive, one cannot simply apply this result to the fluctuation around (conformally) flat spacetime.
On the other hand, we have seen a concordance of the chirality generation between the Bianchi type-IX case and the homogeneous SU(2) gauge field case.
This suggests that a deeper understanding of the chirality generation under the external gauge field may help us to understand that generated
by the parity-violating gravitational field.
With this spirit, in the following section we discuss the fermion chirality generation under chiral GWs around flat spacetime based on the analogy between the classical electromagnetism and the weak gravitational fields.

\section{Investigation of the Dirac equation under the parity-violating weak spin-2 field}\label{sec:weak_spin2}
In this section, we discuss the possibility of level crossing in the gravitatinoal leptogenesis.
For this purpose, we introduce a toy model for the parity-violating spin-2 gravitational background that captures the characteristics of configuration considered in the gravitational leptogenesis and yields $R\tilde{R} \neq 0$. 
The model is actually an extension of the U(1) gauge field configuration discussed in Sec.~\ref{sec:u1} and turned out to be suitable for examining the gravitational leptogenesis (see App.~\ref{app:GL_min_config} for similarity between our model and the one in gravitational leptogenesis).
This system is, however, hard to solve analytically in contrast to the U(1) case due to its inhomogeneous and anisotropic nature.
Therefore, to obtain the physical insights on the chirality production, we tackle this system by applying the analogy between the classical electromagnetism and weak gravity. 
We find that the spin-2 nature of the gravity seems to make the level crossing less efficient than in the U(1) case.

\subsection{Dirac equation under the weak gravitational field}
We first investigate the field equation for the massless
Dirac fermion in the weak gravitational field background to see its ``Abelian''-like nature.
Here we consider the metric perturbation around the Minkowski spacetime: $g_{\mu\nu} = \eta_{\mu\nu} + h_{\mu\nu}$.
Throughout this section, we impose the TT gauge condition to pick up the spin-2 contribution~\footnote{We can also consider, for example, the spin-1 contribution, which may be related to the chiral vortical effect~\cite{Vilenkin:1979ui,Son:2009tf,Landsteiner:2011cp}, and perform a similar discussion, but it is not the physical degree of freedom that plays the role in the gravitational leptogenesis.}
and consider up to the linear order.
The massless Dirac equation in the curved spacetime is given as~\cite{Birrell:1982ix,Parker:2009uva}
\beq
\mathrm{i}\gamma^a e^{\mu}_{\ a} D_{\mu}\psi = 0, \quad D_{\mu} \equiv \partial_{\mu} -\frac{\mathrm{i}}{4}\omega_{\mu ab}\sigma^{ab}, \label{Dirac_curve}
\eeq
where $e^\mu_{\ a}$ is the tetrad that satisfies $g^{\mu\nu} = e^\mu_{\ a} e^\nu_{\ b} \eta^{ab}$, $\omega_{\mu}^{\ ab} = e_{\nu}^{\ a}\Gamma^{\nu}_{\ \sigma\mu}e^{\sigma b} + e_{\nu}^{\ a}\partial_{\mu}e^{\nu b}$ ($\Gamma^{\nu}_{\ \sigma\mu}$ is the Christoffel symbol) is the spin connection,
and $\sigma^{ab} = (\mathrm{i}/2) [\gamma^a, \gamma^b]$.
The tetrad and the spin connections around the Minkowski spacetime are  expanded in terms of $h_{ij}$ (see Eq.~\eqref{eq:inf_tensor_pert}) as
\begin{equation}
    e^i_{\ a} \simeq \delta^i_{\ a} - \frac{1}{2} \eta^{i k} h_{k a} + \mathcal{O}(h^2), \quad \omega_{\mu ab} = \frac{1}{2}(\partial_b h_{\mu a} - \partial_a h_{\mu b}), \label{eq:DiracEqGWkspace}
\end{equation}
such that the Dirac equation can be expanded as
\beq
\lnk \mathrm{i}\partial_t \pm \mathrm{i} \lmk\delta^j_{\ i} -\frac{1}{2}h^j_{\ i}\rmk\sigma^i\partial_j\rnk\psi_{\rm R/L}(t, \bm{x}) = 0. \label{linear}
\eeq
From the TT gauge condition we find
$\omega_{0ab}=0$ and $\gamma^i \omega_{i ab}\sigma^{ab} = \gamma^i \Gamma_i= 0$ at the linear order in $h_{ij}$, and hence $h_{ij}$ in Eq.~\eqref{linear} comes solely from the tetrad in Eq.\eqref{Dirac_curve}. 
This is quite in contrast to the small GW amplitude limit $\beta_+ \ll 1$ of Bianchi type-IX case where $\gamma^i \Gamma_i \neq 0$ at the leading order of $\beta_+$ (or $h_{ij}$) and the energy bias still appears (even for $\beta_+=0$).
This is because the Bianchi type-IX spacetime is the deformation of an already-curved spacetime (more specifically, a closed sphere), where the spin connection yields non-vanishing $\gamma^i\Gamma_i$.
While this bias term is important for the particle production (as well as the emergence of the vacuum contribution) in the Bianchi type-IX case, the gravitational leptogenesis occurs in the system with the parity-violating GWs 
around the conformally flat spacetime. 
Therefore, to investigate how the chirality production takes 
place in gravitational leptogenesis, we cannot use the result 
in the Bianchi type-IX spacetime and we need to examine the Dirac equation~\eqref{linear} under some appropriate metric perturbation.

By using the plane wave ansatz formally,
\beq
\psi_{\rm R/L}(t, \bm{x}) = e^{i\bm{p}\cdot\bm{x}}{\tilde \psi}_{\rm R/L}(t, \bm{x}; \bm{p}),
\eeq
where ${\tilde \psi}_{\rm R/L}(t, \bm{x}; \bm{p})$ does not depend explicitly on $\bm{x}$, the equation of motion becomes
\beq
\lnk \mathrm{i}\partial_t \mp \lmk \bm{\sigma}\cdot\bm{p}\rmk\pm\frac{1}{2}h_{ij}(t,\bm{x})\sigma_ip_j\rnk {\tilde \psi}_{\rm R/L}(t, \bm{x}; \bm{p}) = 0.\label{eq:eom_fourier}
\eeq
Note that $p_i$ is the ``momentum'' associated with the background flat spacetime $\eta_{\mu\nu}$, but not the quantity defined for the full spacetime $g_{\mu\nu} (=\eta_{\mu\nu}+h_{\mu\nu})$. 
This is contrasted by the case of Bianchi type-IX spacetime preserving homogeneity as we can take the invariant basis on which $h_{ij}$ becomes constant and easily find the momentum globally defined.
As usually done in the study of cosmological perturbation theory, we use this flat spacetime momentum and regard the last term as an interaction between the ``external field'' $h_{ij}$ and the Dirac field.
One can formally identify the external $h_{ij}$ field with a U(1) gauge field as
\beq
A_j(t,\bm{x},\bm{p}) \equiv (1/2)h_{ij}p_i, \label{eq:graviu1field}
\eeq
which the fermion with momentum $p_i$ feels. The inhomogeneous and anisotropic nature of $h_{ij}(t,{\bm x})$, which will be discussed below, is now encoded in $A_j(t,\bm{x},\bm{p})$.
Here $A_j$ satisfies the Lorentz gauge condition due to the TT gauge condition and hereafter we shall refer to it as geometric gauge field.
From this similarity between the weak gravitational field and U(1) gauge field, we can expect that the chirality generation in the minimal gravitational leptogenesis might be understood as the ``Landau'' level crossing due to the ``electric'' force.

\subsection{A toy field configuration for the gravitational leptogenesis}
Next we construct a parity-violating configuration of weak gravitational fields, which would be suitable to investigate the situation of our interest.
The construction can be done in a similar fashion to the helical U(1) gauge fields studied in Sec.~\ref{sec:u1}, by making an analogy to the classical electromagnetism and weak gravity. 

Under the weak gravity $|h_{\mu\nu}| \ll |\eta_{\mu\nu}|$,
the Riemann curvature tensor is linearized as
\beq
R^{\tau}_{\ \sigma\mu\nu} \simeq \partial_{\mu}\Gamma^{\tau}_{\nu\sigma} 
-\partial_{\nu}\Gamma^{\tau}_{\mu\sigma}
\simeq 
\partial_{\mu}\lnk\frac{1}{2}\lmk \partial_{\sigma}h_{\nu}^{\ \tau} - \partial^{\tau}h_{\nu\sigma}\rmk\rnk
-
\partial_{\nu}\lnk\frac{1}{2}\lmk \partial_{\sigma}h_{\mu}^{\ \tau} - \partial^{\tau}h_{\mu\sigma}\rmk\rnk
.\label{eq:rieman_pert}
\eeq
With the naive identification of $\Gamma_{\mu\sigma}^{\tau}$ with a gauge field, one can see the similar structure between the linearized Riemann curvature tensor and the field strength of Abelian gauge field~\eqref{eq:FS}.
This similarity in the structure of electromagnetism and gravity has been widely investigated in the field of ``gravito-electromagnetism'' (GEM).
The interested readers can refer to, for example, Ref.~\cite{Mashhoon:2003ax} as a comprehensive review of GEM.
This correspondence motivates us to define the ``electric-like" component and the ``magnetic-like" component of the curvature tensor as follows:
\begin{align}
    E_{i\sigma\tau} &\equiv R_{i0\sigma\tau} \simeq \frac{1}{2}\lmk \partial_{\sigma}\dot{h}_{i\tau} - \partial_{\tau}\dot{h}_{i\sigma}\rmk,\\
    B^{i}_{\ \sigma\tau} &\equiv \frac{1}{2}\frac{\epsilon^{0ijk}}{\sqrt{-g}}R_{jk\sigma\tau} \simeq \frac{1}{2}\epsilon^{ijk}\partial_j\lmk \partial_{\tau}h_{k\sigma} - \partial_{\sigma}h_{k\tau}\rmk.
\end{align}
With these components, we can express the Chern-Pontryagin density as
\beq
R\tilde{R} = -4E_{i\sigma\tau}B^{i\sigma\tau},\label{eq:RR_EB}
\eeq
which is a similar expression with the decomposition of $F\tilde{F}$ in Eq.~\eqref{eq:FFdual}.
Under the TT gauge condition, $E_{ij0}$ and $E_{ijk}$, $B_{ij0}$ and $B_{ijk}$ generically become non-vanishing.

It is known that for GWs around (conformally) flat spacetime, $R\tilde{R}$ quantifies the asymmetry between the left and right circular polarizations. Particularly, it measures the growth in deviation from the freely propagating solution for each polarization as $\partial_t\lmk|\dot{h}_{\rm R/L}(t)|^2 - k^2|h_{\rm R/L}(t)|^2\rmk + \mathcal{O}(h^4)$, where $h_{\rm R/L}(t)$ abstractly represents the amplitude of the mode with wave number $k$.
This means that when $|\dot{h}_{\rm R/L}(t)|^2$ has non-trivial contribution in the left-right asymmetric manner, which is the case in the gravitational leptogenesis,
$R\tilde{R}$ becomes non-vanishing and the fermion chirality is produced. 
In our electromagnetic notation, it is straightforward to verify that $E_{ij0}B^{ij0}$ corresponds to $\partial_t |\dot{h}_{\rm R/L}(t)|^2$ contribution in $R\tilde{R}$.
Hence, we hereafter focus on the configuration of $h_{ij}$ with which only $E_{ij0}$ and $B_{ij0}$ are non-vanishing. While simplifying the system, this allows us to capture the essential contribution in gravitational leptogenesis.

In a similar way to the helical U(1) gauge field studied in Sec.~\ref{sec:u1}, we can construct a parity-violating spin-2 configuration which gives a ``simple'' expression of $R\tilde{R}$ from the constant and non-vanishing $E_{ij0}$ and $B_{ij0}$.
In terms of $h_{ij}$ in the TT gauge, $E_{ij0}$ and $B_{ij0}$ at the linear order are expressed as\footnote{These tensors are nothing but the gravito-electromagnetic tidal tensors, $\mathbf{E}_{ij} \equiv R_{i\alpha j\beta}u^{\alpha}u^{\beta}$ and $\mathbf{B}_{ij} \equiv (\epsilon_{\alpha i}^{\ \ \rho\sigma}/2\sqrt{-g})R_{\rho\sigma j\beta}u^{\alpha}u^{\beta}$, for the static observer $u^{\alpha} = \delta^{\alpha}_0$~\cite{FilipeCosta:2006fz}. 
Note that the correspondence between the gravitational fields described with gravito-electromagnetic tidal tensors and the electromagnetic fields becomes clearer when thinking of geodesic deviation caused by the latter (see Appendix.~A of Ref.~\cite{FilipeCosta:2006fz}).}
\begin{align}
E_{ij0} &= -\frac{1}{2}\ddot{h}_{ij},\\
B^{i}_{\ j0} &= \frac{1}{2}\epsilon^{ilm}\partial_l\dot{h}_{mj}.
\end{align}
From these expressions, one can see that formally, $\mathcal{A}_{ij} \equiv (1/2)\dot{h}_{ij}$ is quite similar to the vector potential $A_{\mu}$.
By comparing to the configuration $A_{\mu} = (0, 0, Bx, Et)$ resulting in a constant $F\tilde{F} = 4EB$ (see Eq.~\eqref{U1homoconfig}), we find that
\beq
\mathcal{A}_{ij} = 
\left(
\begin{array}{ccc}
Et & -Bz & 0\\
-Bz & -Et & 0\\
0 & 0 & 0
\end{array}
\right)
\label{eq:grav_tensor_pot}
\eeq
gives a constant and parallel $E$ and $B$ tensor in $x$-$y$ direction as $B_{110} = -B_{220} = B$ and $E_{110} = -E_{220} = -E$ at the first order in the perturbation, leading to $R\tilde{R} \simeq 16EB$ and the chirality generation. 
Here we shall restrict ourselves to small intervals of $t$ and $z$ for perturbativity. 
The simplicity of configuration could provide an opportunity to gain clear insight into how parity-violating spin-2 fields around a flat spacetime act on fermions, similarly to the U(1) case studied in Ref.~\cite{Domcke:2018eki}.
As we will see below, however, inhomogeneous and anisotropic nature of $h_{ij}$ leads to a complication in the field equation.

Physically, the configuration of our $h_{ij}$ where $x$-$y$ direction is non-vanishing resembles GW on a flat Minkowski background running in the $z$ direction. 
As summarized in App.~\ref{app:GL_min_config}, we find that constant and aligned $E_{ij0}$ and $B_{ij0}$ resembles the one-handed single $k$-mode of GWs in the minimal gravitational leptogenesis.
Most importantly, each momentum mode of chiral GWs in the miminal model of gravitational leptogenesis~\cite{Alexander:2004us} experiences the left-right asymmetric growth of amplitude quadratic in (conformal) time as in Eq.~\eqref{eq:GL_minimal_mode}. 
In terms of the tensor potential $\mathcal{A}_{ij}$, this corresponds to the growth linear in time as in Eq.~\eqref{eq:GL_min_pot}, similarly to that of our toy model~\eqref{eq:grav_tensor_pot} (or see Eq.~\eqref{eq:toy_RL} in the circular polarization basis for a more direct comparison).
Notice that since there is no mode-mixing in the right-hand side of Eq.~\eqref{eq:GL_lepton}, it is enough to focus on a single momentum and polarization mode to examine the chirality generation mechanism.

While this makes our metric suitable for investigating fermion production in this context, there also exists a qualitative difference. For example, while $E_{ijk}$ has non-vanishing component such as $E_{123} = B$, $B_{ijk} = 0$ in the present case and they do not contribute to the anomaly equation.
This is not the case for the real chiral GWs which yields non-vanishing $\partial_t\lmk k^2|h_{\rm R/L}(t)|^2\rmk$ contribution in $R\tilde{R}$ through $E^{ijk}B_{ijk}$.
Nevertheless, as discussed below Eq.~\eqref{eq:RR_EB}, the deviation from the freely propagating solution encoded in $\partial_t\lmk|\dot{h}_{\rm R/L}(t)|^2\rmk$ is crucial for chiral GWs in generating non-zero $R\tilde{R}$.
Indeed, our toy model resembles such deviation in the minimal gravitational leptogenesis as described by Eq.~\eqref{eq:GL_min_pot}.
Therefore, we believe that the gravitational field~\eqref{eq:grav_tensor_pot} with constant $E_{ij0}$ and $B_{ij0}$ still captures more or less the essence of the chiral GWs in the minimal gravitational leptogenesis.

Before closing, we note again that these characteristics cannot be captured by the examples of Bianchi spacetimes, where the non-linear nature of spacetime curvature ($\gamma^i\Gamma_i \neq 0$ at the background level) is essential to have non-zero $R{\tilde R}$ and resultant chirality generation.
We believe that in this sense, the following discussion could serve as a foundation for future studies involving more realistic chiral GW backgrounds.
In the next subsection, we discuss how the geometric U(1) gauge field~\eqref{eq:graviu1field} with the metric~\eqref{eq:grav_tensor_pot} affects the fermion field through the 
Dirac equation.

\subsection{Application of U(1) gauge physics to the weak spin-2 gravitational field}
Now we try to examine the dynamics of fermions on the spacetime with the metric~\eqref{eq:grav_tensor_pot}.
The explicit form of the geometric gauge field~\eqref{eq:graviu1field} for a fermion with momentum $p_i$ is given as
\begin{align}
A_x = \frac{1}{2}Et^2p_x-Btzp_y, \ \ \ A_y = -\frac{1}{2}Et^2p_y-Btzp_x,  \ \ \ A_z = 0. \label{eq:gravu1explicit}
\end{align}
Note that the spacetime dependence of $A_{x/y}$ is the second order to yield non-vanishing $R\tilde{R}$, 
which has one additional derivative compared to $F\tilde{F}$.
This is originated from the difference in the order of derivatives between the Riemann tensor and gauge field strength tensor.
Eq.~\eqref{eq:graviu1field} results in the following ``geometric electromagnetic fields'', $E_i \equiv - \partial_t A_i$, $B^i = \epsilon^{ijk} \partial_j A_k$, as\footnote{This ``electric and magnetic fields'' are related to but different from the gravito-electromagnetic tidal tensors introduced in the previous subsection.}
\begin{equation}
\begin{aligned}
E_x &= -Etp_x+Bzp_y, \ \ \ E_y = Etp_y+Bzp_x, \ \ \ E_z = 0,\\
B_x &= Btp_x, \ \ \ B_y = -Btp_y \ \ \ B_z = 0.
\end{aligned}\label{eq:geoemtric_EB}
\end{equation}
From the Dirac equation~\eqref{eq:DiracEqGWkspace} with the identification~\eqref{eq:graviu1field}, 
a fermion particle with momentum $p_i$ would feel 
the electric force as ${\bm F} ={\bm E} = (-Etp_x-Bzp_y,Etp_y-Bzp_x,0)$
and the Lorentz force with ``magnetic'' field ${\bm B}= (Btp_x, -Btp_y,0)$ (see also Ref.~\cite{Stone:2023qln}).

Nevertheless, one can clearly recognize the difference from the homogeneous and constant electric and magnetic field for the U(1) gauge theory discussed in Sec.~\ref{sec:u1}.
For example, the contribution proportional to $B$ appears in $E_i$, which is related to the fact that there are non-vanishing spatial component in $E_{ijk}$ such as $E_{123} = B$.
This component of electric field is, however, orthogonal to the magnetic field, and the anomaly equation indicates that it does not contribute to chirality generation.
Hence, as for the electric field, we hereafter focus on the contribution that is proportional to $E$.
Much more important difference is that due to the anisotropic and inhomogeneous nature of $h_{ij}$, the magnitude of ``electric and magnetic'' fields depends on the time, space and momentum of the fermions, and they are no longer constant.
Consequently, the direction of the ``electric and magnetic'' fields is not fixed but lies somewhere in the $x$-$y$ plane. 
Note that the time dependence of our ``electric and magnetic'' fields, which
captures the characteristics of the parity-violating growth of the polarized GWs, satisfies the first condition for the vacuum contribution being non-vanishing. That is, the gravitational field is non-equivalent between the initial and final configurations up to the gauge (or general coordinate) transformation.

One may expect that the Dirac equation with the gravitational background~\eqref{eq:gravu1explicit} could be solved in a similar way to the case of U(1) gauge field in Sec.~\ref{sec:u1}.
In fact, this is not possible due to the qualitative difference discussed above.
While the U(1) gauge field there also admits the time and space dependence, its electric and magnetic fields are constant and aligned.
Consequently, we can fortunately find the elegant analytic solution with the help of {\it globally defined} Landau levels, which are physically meaningful quantities. 
In the present case, however, our geometric electric and magnetic fields have the complicated dependence on spacetime and momentum. Therefore, despite the simple expression of $R\tilde{R}$ in Eq.~\eqref{eq:RR_EB}, we do not have Landau levels {\it globally defined} over all the fermion modes.
This, reflecting the essential difference between $F\tilde{F} \propto EB$ and $R\tilde{R}\propto EB$, is the one of the reasons for the difficulty in obtaining the analytic solution.

Instead, we try to read off the physics with the analogy between weak gravitational field and electromagnetism, which would provide us an intuitive understanding of the chirality generation in this system.
Here we apply what is understood for the case of U(1) gauge fields (discussed in Sec.~\ref{sec:u1}) to our present system.
Namely, we speculate the kinematics of fermion field when we first add only the ``magnetic'' field and then turn on the ``electric'' field for a finite time.
In the case of U(1) gauge fields, Landau levels appeared along the direction of the magnetic field, where the LLL connects smoothly the negative and positive energy modes.
Once we turn on the electric field in parallel to the magnetic field, the states in the LLL are accelerated to lead chirality generation.
Assuming that each mode (momentum) of fermion forms Landau levels as if the gravitational ``magnetic'' field is constant,
we examine if the gravitational ``electric'' field can accelerate the fermion in a state at the LLL to cross from the negative to positive energy states.

\begin{figure}[htbp]
\begin{center}
\includegraphics[width=0.5\columnwidth]{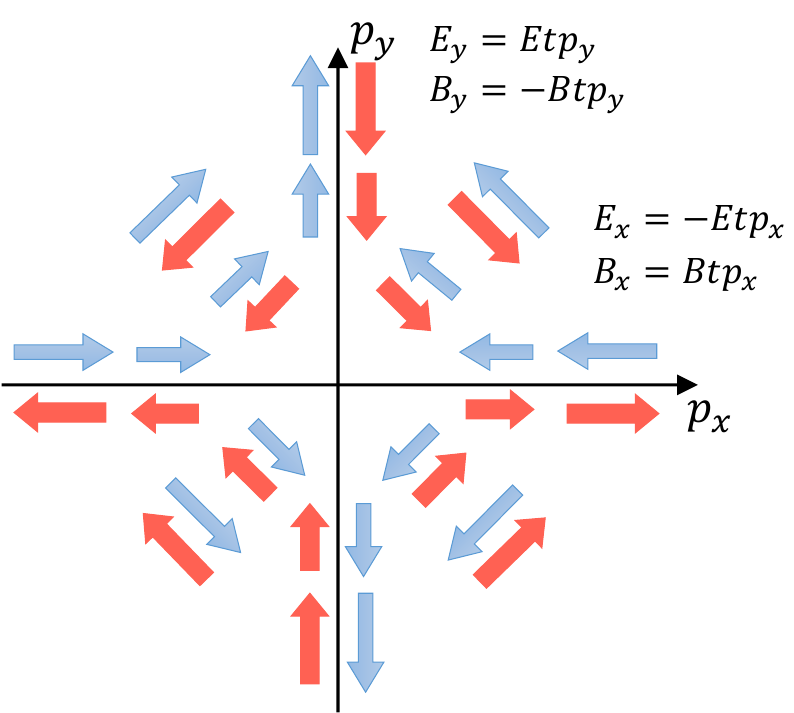}
\caption{\small
Configuration of the geometric electric and magnetic field (with $z = 0$) in the momentum space. 
One can see the configuration becomes the same after $180^{\circ}$ rotation due to the spin-2 nature.}
\label{fig:config}
\end{center}
\end{figure}

\begin{figure}[htbp]
\begin{center}
\includegraphics[width=0.5\columnwidth]{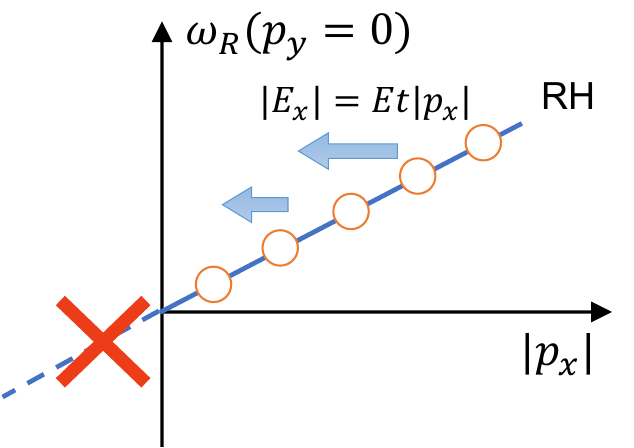}~
\includegraphics[width=0.5\columnwidth]{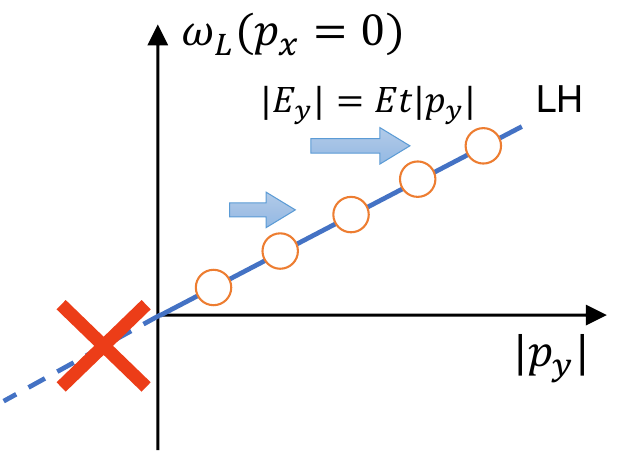}
\includegraphics[width=0.5\columnwidth]{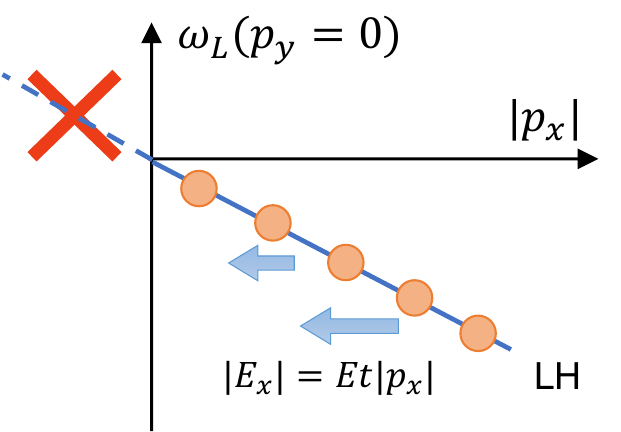}~
\includegraphics[width=0.5\columnwidth]{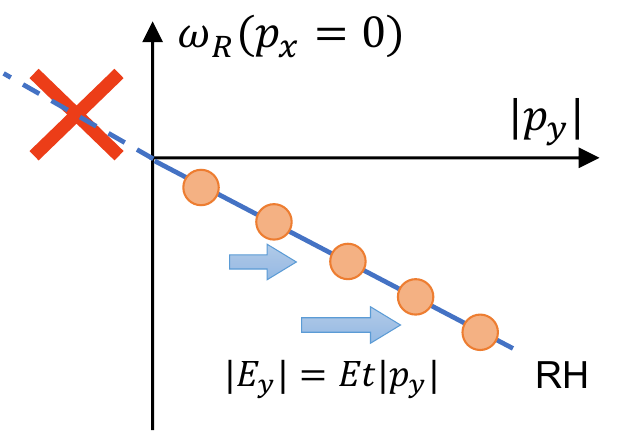}
\caption{\small
Schematic description that even if the LLL-like dispersion exists, positive mode and negative mode are not connected. 
}
\label{fig:EB_LLL-like}
\end{center}
\end{figure}

In Fig.~\ref{fig:config}, the configuration of ``gravitational electric and magnetic field'' of our metric~\eqref{eq:grav_tensor_pot} is shown in the momentum space (in the $p_x$-$p_y$ plane).
Here the spin-2 nature is manifest such that the configuration becomes the same after $180^{\circ}$ rotation. This nature, however, results in the crucial difference from the case of U(1) gauge fields.
Let us focus on the $p_x$ axis at $t \neq 0$ and $z=0$ where the geometric electric field and magnetic field are parallel to those axis, 
\begin{equation}
    B_x = B t p_x, \quad E_x = - Et p_x, \quad B_y=B_z=E_y=E_z=0. 
\end{equation}
In this axis, a simpler ``kinematics'' is expected. 
Since only the LLL participates the physics of the chiral anomaly in the U(1) case, the states we are interested in here would be those with spin aligned with the magnetic fields.
On this axis, the magnetic field continues to point in the opposite direction for $p_x > 0$ and $p_x < 0$, namely, $\mathrm{sgn}(B_x(p_x > 0)) = -\mathrm{sgn}(B_x(p_x < 0)) $.
Therefore one may expect the states similar to the LLL appear in a directionally dependent way.
By naively applying the dispersion of LLLs (with the electric field being turned off), we anticipate the following LLL-like dispersion for these regions:
\begin{equation}
\begin{aligned}
    \omega_{\rm R}(p_x \neq 0, p_y = 0) &\simeq |p_x|,\\
    \omega_{\rm L}(p_x \neq 0, p_y = 0) &\simeq -|p_x|,\\
\end{aligned}\label{eq:LLL-like}
\end{equation}
where the $|p_{x}|$ dependence arises because of the flip of the geometric magnetic field direction at the origin of momentum space due to the spin-2 nature. 
Consequently, we find the LLL-like dispersion does not smoothly connect the negative and positive energy modes. 
Similar discussion holds for the $p_y$ axis.
In Fig.~\ref{fig:EB_LLL-like}, the dispersion~\eqref{eq:LLL-like} is schematically shown. Note that the direction of magnetic field and therefore the dispersion~\eqref{eq:LLL-like} does not change in time, but the Landau degeneracy per unit area, $|{\bm B}|/2\pi$, should change with time and momentum according to Eq.~\eqref{eq:geoemtric_EB}. 

When the electric field is applied to the system for a certain time duration, the LLL allows the excitation continuously to occur in the case of U(1). 
As shown in Fig.~\ref{fig:EB_LLL-like}, however, this spin-2 configuration does not seem to allow continuous level crossing on each axis in the same way as the U(1) example. 
That is, even if the LLL-like dispersion appears, positive and negative frequency mode is not connected because of the $|p_{x/y}|$ dependence, and hence it is unlikely that 
particle production occurs even when we turn on the ``electric'' field.
Although the geometric electric and magnetic fields are always aligned, the spin-2 nature seems to result in this qualitative difference.
Moreover, at $p_i=0$ where $\omega_{\rm R/L} = 0$ and the level crossing could occur, $E_i = 0$ and hence no acceleration occurs.
Having in mind that $h_{ij}$ interacts with fermions analogously to the $U(1)$ gauge field, we expect that the excitation cannot be the principal source for the fermion chirality.

Since we have non-vanishing $R{\tilde R}$, the anomaly equation should be totally satisfied by taking into account the vacuum 
contribution. 
Indeed, the second condition for the appearance of the vacuum contribution, namely, the asymmetric structure in the dispersion relation between the positive and negative energy modes, seems to be satisfied with non-zero $B$ independently for $p_x$ and $p_y$ axis. 
Although the contributions from the $p_x$ and $p_y$ axis looks to cancel each other for vanishing $E$, non-vanishing $E$ causes a bias in the opposite direction depending on the chirality (see Fig.~\ref{fig:EB_LLL-like}), such that the cancellation is broken down. Therefore, we expect that an asymmetry would be accumulated as the vacuum contribution.
Note that we have already seen that the first condition for the appearance of the vacuum contribution (gauge inequivalence of initial and final field configuration) is satisfied. 
Therefore, although we do not confirm that the saturation of anomaly equation with a concrete calculation, we have the suggestion that all (or 
at least most of) the 
chiral charge would be accumulated in the vacuum.
 
Although the argument in the above is a qualitative discussion based on the kinematic analogy, we have obtained the suggestion for the inefficiency of the chirality production through the level crossing
in the case of parity-violating weak gravitational background with the field configuration~\eqref{eq:grav_tensor_pot}.
This behavior is totally different from the homogeneous U(1) case where the Landau level is globally defined and the electric field efficiently excites the field with the homogeneous acceleration. 
The present system is also qualitatively different from the case of Bianchi type-IX spacetime.
In the latter case, in addition to having non-Abelian-like interaction, which causes a bias term in the dispersion, the field configuration preserves the homogeneity of the system.  
We believe that this homogeneity leads to the less efficient but non-negligible amount of chiral particle production through the level crossing.
In contrast, our spacetime identified with the geometric electric and magnetic field is highly inhomogeneous and anisotropic, which, in addition to the spin-2 nature, would be the reason of the inefficiency of the particle production. 

We emphasize that although more realistic field configuration must be considered to conclude, the characteristics of our metric that lead to inefficient level-crossing should apply to the chiral GWs in gravitational leptogenesis. That is, they are nothing but spin-2 perturbations around a conformally flat background and experience the parity-violating amplification quadratic in (conformal-)time.
Moreover, the real configuration is described as a superposition of multiple modes, behaving as an inhomogeneous external field within each horizon patch (while preserving statistical homogeneity). 
As discussed in Sec.~\ref{sec:lesson}, the homogeneity of the external field is expected to play a crucial role in efficient particle excitation. 
Considering the actual field configuration in this sense, it is expected to further suppress the (possible) chiral charge contribution from the excitation.
Therefore, we conjecture that the fermion chirality generated in the gravitational leptogenesis scenario is likely to be accumulated mainly in the vacuum.

We should admit that the above discussion is based on the naive application of the well-studied cases of gauge field, since we were not able to solve the field equation analytically. 
For example, we have not seriously considered the region where $p_x\cdot p_y \neq 0$ and the component of electric field that depends on $Bzp_{x/y}$.
As we mentioned above, however, the latter contribution is hardly expected to be the essential contribution for generating chirality, because it is always orthogonal to the (corresponding) magnetic field (${\bm E}\cdot{\bm B} = 0$). 
We also expect that chiral particle production may not efficiently take place in the region $p_x\cdot p_y \neq 0$ as the geometric magnetic field does not align with the momentum, resulting in low selectivity of chirality. 
We would therefore expect the above inferences to be somewhat on target, but unfortunately they remain inconclusive yet. 
All these expectations could be clarified when, for example, the system is numerically solved. 
Before concluding, for our future reference, let us summarize the practical difficulties to work on the field equation of this system and the main differences from the solvable case of $U(1)$ gauge fields.
\begin{itemize}
    \item Field configuration of $h_{ij}$ with $R\tilde{R} \propto EB$ results in the equation of motion where two directions of momentum are involved ($x/y$ in the above case).

    \item Since the order of derivative in $R\tilde{R}$ is different from that in $F\tilde{F}$ (second-order for the former while first-order for the latter), the geometric ``electric'' and ``magnetic''
    fields in Eq.\eqref{eq:geoemtric_EB} have explicit space time dependence.
    
    \item The interaction between graviton and fermion non-trivially depends on the ``momentum'' of fermion field. This makes it difficult to solve the Dirac equation with the usual mode expansion of the fermion field if the globally defined momentum is absent.

\end{itemize}

\subsection{Implication to the gravitational leptogenesis}
Finally, let us comment on the feasibility of gravitational leptogenesis scenario in the light of our discussion above. 
As discussed in Sec.~\ref{sec:GL_review}, chiral charge carried by the left-handed neutrinos, which is generated under chiral GWs during inflation, provides non-zero lepton number in the early universe. If sufficiently large lepton number is produced and converted into the baryon number, the observed baryon asymmetry could be explained by this anomalous contribution.
As we have seen, however, it seems difficult for the parity-violating spin-2 gravitational field (including GWs) around the spatially-flat FLRW universe to cause level-crossing.
Therefore, we have conjectured that the lepton number generated in this scenario (Eq.~\eqref{eq:GL_lepton}) may almost coincide with the eta invariant and accumulate in the vacuum, which does not count the number of left-handed neutrinos as excited states.

If this is the case, one may wonder whether the transport relation~\eqref{eq:sphaleron} holds for the vacuum contribution. 
The relation~\eqref{eq:sphaleron} is derived under the framework of kinetic theory, which deals with the non-perturbative electroweak sphaleron process as well as the many body scattering in {\it thermalized} system.
Therefore, it is questionable whether the lepton number accumulated in the vacuum can be converted into the baryon number in the thermal plasma according to Eq.~\eqref{eq:sphaleron} and contribute to the thermal history of the Universe such as the Big Bang Nucleosynthesis. 
If the lepton number accumulated in the vacuum, which may account for the substantial amount of net lepton number~\eqref{eq:GL_lepton}, completely decouples from this transport, resultant baryon-to-entropy ratio in this scenario could be significantly smaller than expected. 
Nevertheless, we cannot rule out the possibility that the ``lepton-charged'' vacuum induces a bias for the sphaleron process to convert the vacuum lepton charge 
to the baryon number in the plasma.
We leave the investigation whether the vacuum contribution becomes relevant for the baryon number conversion to future work.

\section{Summary and Discussion} \label{sec:discussion}
In this paper, we investigated the chirality production of fermion under parity-violating spin-2 gravitational fields, which is similar to those studied in the gravitational leptogenesis~\cite{Alexander:2004us}.
Our discussion is based on the existing studies on the cases with the external gauge fields (e.g., Refs.~\cite{Nielsen:1983rb,Domcke:2018eki,Domcke:2018gfr}).
Chirality of fermion generally consists of the contribution from excitation and that accumulated in vacuum~\cite{Atiyah:1963zz,Atiyah:1968mp}. 
The distinction of these contributions, however, was never addressed in the context of gravitational leptogenesis, where lepton number is produced by chiral GWs during inflation according to the gravitational chiral anomaly~\cite{Kimura:1969iwz,Delbourgo:1972xb,Eguchi:1976db,AlvarezGaume:1983ig}.
On the other hand, in addition to the old study on the Bianchi type-IX spacetime~\cite{Gibbons:1979kq,Gibbons:1979ks}, the dominance of vacuum contribution was recently reported for the Bianchi type-II spacetime~\cite{Stone:2023qln}, both of which study parity-violating deformation of metric in the already-curved background.
In this situation, it is worth investigating which contribution becomes dominant when the parity-violating spin-2 gravitational field is imposed around spatially flat spacetime.
Such a clarification may help refining the prediction in the scenarios of gravitational leptogenesis.

We first made a review on the chirality generation under the gauge fields. While the smooth excitation in the LLL accounts for all the fermion chirality in the U(1) case, vacuum contribution known as eta invariant becomes important in the SU(2) case.
The most distinct difference is that non-Abelian ``magnetic field'' causes overall bias in the dispersion similarly to the chemical potential, which makes the energy spectrum highly asymmetric, leading to the dominance of vacuum contribution in the chiral charge.
As discussed in Sec.~\ref{sec:bianchi}, this result is in fact helpful in understanding the generation of fermion chirality in the Bianchi type-IX spacetime, since it shares the similar characteristics with the homogeneous SU(2) gauge field case.
To the best of our knowledge, this is the first discussion of this system from such a perspective, and the section can be considered as a comprehensive review, including an explicit decomposition into closed spheres and chiral GWs and confirmation of the APS index theorem.
On the other hand, in both SU(2) and Bianchi type-IX cases, excited contribution in chirality is small but always non-negligible even in the strong external field limit.
We conjecture that this behavior is provided by the homogeneity of the external field as it seems to allow continuous excitation.

In Sec.~\ref{sec:weak_spin2}, we introduced a specific configuration of metric perturbation around the flat spacetime.
This has parity-violating spin-2 polarization and yields non-vanishing topological charge of the spacetime, similarly to the aligned constant electric and magnetic field.
With the classical analogy between the electromagnetism and weak gravitational field, we anticipate the physics of Landau levels in U(1) gauge field can be applied to the weak spin-2 gravitational field.
The configuration in particular captures the characteristics of the amplitude growth of chiral GWs in the gravitational leptogeneis, which is the essential ingredient of the scenario.
It is therefore appropriate for our purpose to discuss the chirality generation in the context of gravitational leptogenesis.
However, as the order of differentiation in the curvature tensor differs, the metric perturbation admits non-trivial spacetime dependence in the Dirac equation and acts as an anisotropic and inhomogeneous ``electric and magnetic field''. 
As a result, it becomes difficult to obtain analytical solution and physical understanding of chirality generation.  
Thus we tried to investigate the system with brave approximations. That is, we assume that the system is described by the Landau levels induced by the geometric ``magnetic'' fields along which the fermions in the levels are accelerated by the geometric ``electric'' fields, in a similar way to the case of U(1) gauge field.
From this approach, we find that the spin-2 nature of the gravitational field seems to prevent the efficient excitation.
As the geometrical ``magnetic field'' changes the direction at the origin of the momentum space, smooth excitation could not take place in the ``possible'' lowest Landau level-like dispersion.
If this investigation applies for the gravitational leptogenesis, 
the baryon number generation becomes less efficient in the sceneario,  
since the charge transfer through the electroweak sphalerons to baryon 
charge is non-trivial for the lepton charge accumulated in the vacuum.

It would be idealistic to solve the Dirac equation to analyze the level crossing of fermions in the realistic, dynamically evolving  gravitational background within the context of gravitational leptogenesis, or at least with the field configuration~\eqref{eq:grav_tensor_pot}. 
However, as discussed in the main text, we found it challenging to reach the definitive conclusion in a single step. 
This paper aims to highlight the potential inefficiency of level-crossing in the context of gravitational leptogenesis and to provide supportive evidence.
Proving or falsifying this rigorously will require a systematic, step-by-step approach.
With this perspective in mind, we identify
several directions in our future work. 
One direction is a numerical approach to the field equations we have formulated. 
Instead of trying to solve the field equation directly, one could work with the (Euclidean) Dirac operator to rigorously show the absence of a normalizable zero mode {\it e.g.} based on the symmetry of the system.
This approach may require exploring configurations beyond the one we have considered~\eqref{eq:grav_tensor_pot}, 
which could also offer deeper insights into the effects of parity-violating gravitational fields.
Additionally, we could think of alternative formulation of the system such as the chiral kinetic equation in curved spacetime~\cite{Liu:2018xip,Hayata:2020sqz}.
This framework naturally involves the extension of the momentum space and could be useful for our case where the momentum is regarded as the local quantity.
If exist, it would also be worth investigating whether the vacuum contribution of the induced energy momentum tensor can be renormalized, similar to the studies in SU(2) gauge field case~\cite{Domcke:2019qmm}.
Apart from the gravity, it is also worth addressing as a general question how chiral charges accumulated in a vacuum contribute to the transport equation as we mentioned in the last part.

\section*{Acknowledgments}
The authors would like to thank Valerie Domcke,
Yohei Ema,
Hidenori Fukaya, Kyohei Mukaida, Kai Schmitz, and Mikhail Shaposhnikov for useful comments. 
KK is supported by the National Natural Science Foundation of China (NSFC) under Grant No. 12347103 and the JSPS KAKENHI Grant-in-Aid for Challenging Research (Exploratory) JP23K17687.
JK is supported by the JSPS Overseas Research Fellowships.

\appendix
\section{U(1) gauge field on $S^1$}\label{sec:u1_s1}
In this appendix, we show the appearance of the eta invariant in the U(1) theory on a compact manifold with a concrete computation for instruction. 
As we have seen in Sec.~\ref{sec:u1}, the level-crossing in the LLL accounts fully for the chiral anomaly if the magnetic field is inifinitely long. However, even in the U(1) theory, the eta invariant can be non-zero, for instance, when the space is compactified on an $S^1$  along the magnetic flux with the periodicity $L$. In this setup with $B_z=B$ and $E_z=E$, or $A_\mu= (0,0,Bx,-Et)$, the state is discretized with an interval of $\Delta \Pi_z = 2\pi/L$, such that
the LLL of the right-handed fermion is expressed as
\begin{equation}
    \omega_0^{\rm R}(l) = \frac{2\pi l}{L} + e A_z = \frac{2\pi l}{L} - e Et, \quad l= \cdots, -2,-1,0,1,2, \cdots.  
\end{equation}
Unless $t = 2 n \pi /e E L$, when the Wilson loop $\oint dz eA_z$ is trivial,  the distribution of the eigenstates becomes asymmetric between the positive and negative energy states, indicating the presence of a non-zero eta invariant. 
We can explicitly compute the eta invariant (for $0<t<2\pi/eEL$ when the level crossing has not yet occurred) using methods such as $\zeta$ function regularization, yielding
\begin{align}
\eta_H^{\rm R} &= \lim_{s \rightarrow 0} \left( \sum_{l=1}^\infty  \left| \frac{2\pi l}{L} - e Et\right|^{-s} -  \sum_{l=0}^\infty  \left| -\frac{2\pi l}{L} - e Et\right|^{-s} \right) \notag \\
&= \lim_{s \rightarrow 0}   \left( \frac{2\pi}{L}\right)^{-s} \left( \sum_{l=0}^\infty \left|l-\frac{eELt}{2\pi}\right|^{-s} -  \sum_{l=0}^\infty\left|l+\frac{eELt}{2\pi}\right|^{-s} - \left|\frac{eELt}{2\pi} \right|^{-s} \right) \notag \\
&= \lim_{s \rightarrow 0}  \left( \frac{2\pi}{L}\right)^{-s} \left(\zeta\left(s,  -\frac{eELt}{2\pi}\right) -\zeta\left(s,  \frac{eELt}{2\pi}\right) - \left|\frac{eELt}{2\pi} \right|^{-s}\right) \notag \\
&= \frac{eELt}{\pi} -1,
\end{align}
where $\zeta(s,q)$ is the Hurwitz zeta function, and we have used
$\zeta(0,q) = 1/2-q$. 
Similarly, for left-handed fermions we obtain
\begin{equation}
    \eta_H^{\rm L} = -\frac{eELt}{\pi} -1. 
\end{equation}
By considering once more the Landau degeneracy, $\int dx dy e B /2\pi,$ we find the following equation holds,
\begin{equation}
   - \frac{e^2}{8\pi^2} \int d^4 x F_{\mu\nu} {\tilde F}^{\mu\nu} = \frac{e^2}{2\pi^2} \int_0^t dt \int_0^L dz \int dx dy EB = \frac{1}{2}(\eta_H^{\rm R}-\eta_H^{\rm L})  \int dx dy \frac{e B}{2\pi} =\Delta Q_5^{(\mathrm{v})}, 
\end{equation}
indicating that the anomaly equation is fulfilled by the vacuum contribution. 
Indeed, the number of excited states must be an integer while $\int d^4x F_{\mu\nu} {\tilde F}^{\mu\nu}$ is a real number, but not necessarily an integer.

\section{Field configuration in the minimal gravitational leptogenesis}\label{app:GL_min_config}
Here we summarize the qualitative behavior of the chiral GWs considered in the minimal model of gravitational leptogenesis~\cite{Alexander:2004us} and compare it with the toy field configuration studied in Sec.~\ref{sec:weak_spin2}.

The inflationary tensor perturbation in Eq.~\eqref{eq:inf_tensor_pert} can be expanded in the momentum space as
\begin{equation}
    h_{ij}(\eta, \bm{x}) = \frac{1}{(2\pi)^{3/2}}\int d^3k\sum_{\rm s =R,L}p^s_{ij}({\bm k})h_{\bm k}^s(\eta)e^{i{\bm k}\cdot {\bm x}},
\end{equation}
where $p^s_{ij}({\bm k})$ is the circularly polarized basis. The gravitational Chern-Simons coupling $\phi R\tilde{R}$, where $\phi$ is the pseudo-scalar inflaton, causes parity-violating evolution of tensor vacuum fluctuation, whose strength is characterized by a dimensionless parameter $\Theta \propto \dot{\phi}$.
We note that $\Theta (\ll 1)$ is nearly constant during inflation and $k\Theta\eta \ll 1$ needs to be satisfied for the mode of our interest to avoid the ghost instability~\cite{Alexander:2004wk,Kamada:2019ewe}.
At the leading order in $\Theta$, the physical solution of $h_{\bm k}^s(\eta)$ is given as~\cite{Kamada:2020jaf} 
\begin{equation}
\begin{aligned}
    h_{\bm k}^s(\eta) &\sim \frac{H\eta}{M_{\rm Pl}}\left(1 - \lambda_{\bm k}^s
    k\frac{\Theta}{8}\eta \right)^{-1/2}e^{-\lambda_{\bm k}^s \frac{\pi}{32}\Theta}\frac{1}{\sqrt{k}}e^{-ik(\eta- \eta_i)} \\
    & \sim \frac{H\eta}{M_{\rm Pl}}\left(1 - \lambda_{\bm k}^s \frac{\pi}{32}\Theta + \lambda_{\bm k}^s k\frac{\Theta}{16}\eta \right) \frac{1}{\sqrt{k}}e^{-ik(\eta- \eta_i)},\label{eq:GL_minimal_mode}
\end{aligned}
\end{equation}
where the constant $\lambda_{\bm k}^s (= -\lambda_{-{\bm k}}^s)$ characterizes the chiral asymmetry and without loss of generality, we can set $\lambda_{\bm k}^{\rm R} = 1$ and $\lambda_{\bm k}^{\rm L} = -1$ for $k_z > 0$.
From Eq.~\eqref{eq:GL_lepton} we would expect the generation of lepton asymmetry proportional to $\Theta$. 

In order to compare this chiral GW~\eqref{eq:GL_minimal_mode} with our toy model of the gravitational field (see Eq.~\eqref{eq:grav_tensor_pot}), here we investigate the tensor potential $\mathcal{A}^{\rm GL}_{ij} = (1/2)\partial_{\eta}h_{ij}$ sourced by the gravitational field in Eq.~\eqref{eq:GL_minimal_mode}.
Focusing on a single right-handed mode with $\bm{k} = (0, 0, k)$, the tensor potential can be expanded as
\begin{equation}
    \mathcal{A}^{\rm GL}_{ij} \sim
    \frac{1}{2}\lnk
    (\partial_{\eta}h_{\bm k}^{\rm R})
    p^{\rm R}_{ij}e^{ikz} +
   (\partial_{\eta} h_{{\bm k}}^{\rm R*})
    p^{\rm R*}_{ij}e^{-ikz}
    \rnk,
\end{equation}
where the circularly polarized basis $p^s_{ij}$ for this mode is given as
\beq
p^{\rm R}_{ij} = 
\frac{1}{\sqrt{2}}
\left(
\begin{array}{ccc}
1 & i & 0\\
i & -1 & 0\\
0 & 0 & 0
\end{array}
\right)
, \quad
p^{\rm L}_{ij} = p^{\rm R*}_{ij} =
\frac{1}{\sqrt{2}}
\left(
\begin{array}{ccc}
1 & -i & 0\\
-i & -1 & 0\\
0 & 0 & 0
\end{array}
\right).
\label{eq:cp_basis_for_kz}
\eeq
Similarly to Sec.~\ref{sec:weak_spin2}, we assume the small separation in the spatial coordinate such that we take $kz \ll 1$ ($z=0$ is taken as the reference).
Then, up to the irrelevant complex phase, we find that the leading contributions in $\mathcal{A}^{\rm GL}_{ij}$, which give rise to non-vanishing $R\tilde{R}$\footnote{For example, we omitted the contribution comes from $(\partial_\eta) e^{-i k \eta}$ since it does not contribute to $R{\tilde R}$.}, can be expanded as
\begin{equation}
    \mathcal{A}^{\rm GL}_{ij} \sim 
    \sqrt{k}
    \frac{H}{2M_{\rm Pl}}
    \lnk 
    \left(\frac{\Theta}{8}\eta + iz\right)
    p^{\rm R}_{ij} +
    \left(\frac{\Theta}{8}\eta - iz\right)
    p^{\rm R*}_{ij}\rnk,\label{eq:GL_min_pot}
\end{equation}
from which the electromagnetic tensors are evaluated as
\begin{equation}
    E^{\rm GL}_{ij0} \sim 
    -\partial_\eta \mathcal{A}_{ij}^{\rm GL}\sim
    -\sqrt{k}
    \frac{H}{2M_{\rm Pl}}
    \frac{\Theta}{8}
    \lmk p^{\rm R}_{ij} + p^{\rm R*}_{ij} \rmk,\label{eq:GL_min_E}
\end{equation}
\begin{equation}
    B^{\rm GL}_{ij0} \sim 
    \epsilon^{ilm}
    \partial_l\mathcal{A}_{mj}^{\rm GL}
    \sim
    \sqrt{k}
    \frac{H}{2M_{\rm Pl}}
    \lmk
    p^{\rm R}_{ij} +
    p^{\rm R*}_{ij}
    \rmk.\label{eq:GL_min_B}
\end{equation}
Thus, we would write Eq.~\eqref{eq:GL_min_pot} as
\begin{align}
    &\mathcal{A}^{\rm GL}_{ij} \sim (E^{\rm GL} \eta+ iB^{\rm GL}z)p^{\rm R}_{ij} + (E^{\rm GL} \eta - iB^{\rm GL}z)p^{\rm R*}_{ij}, \label{eq:GL_min_pot_EB}\\
    &\text{with} \quad E^{\rm GL} = -\sqrt{k} \frac{H}{2M_{\rm Pl}}
    \frac{\Theta}{8}, \quad B^{\rm GL}= \sqrt{k}
    \frac{H}{2M_{\rm Pl}}. 
\end{align}
Notice that the non-vanishing component of $\sum_{ij}E^{\rm GL}_{ij0}B^{\rm GL}_{ij0}$ corresponds to the first term in the bracket in Eq.~\eqref{eq:GL_lepton}.
Thanks to the absence of mode mixing in the anomaly equation~\eqref{eq:GL_lepton}, the result of lepton charge estimation in Ref.~\cite{Alexander:2004us} would be correctly reproduced by integrating over the whole momentum and adding the opposite chirality (which results in an overall minus sign in both Eq.~\eqref{eq:GL_min_E} and~\eqref{eq:GL_min_B} and thus contributes equally to the RR component).

Now we rewrite the tensor potential of our model~\eqref{eq:grav_tensor_pot} in terms of the circularly polarized basis~\eqref{eq:cp_basis_for_kz}:
\beq
\mathcal{A}_{ij} = \frac{1}{\sqrt{2}}
\lnk
(Et + iBz)p^{\rm R}_{ij} + (Et - iBz)p^{\rm R*}_{ij}
\rnk.\label{eq:toy_RL}
\eeq
One can clearly recognize the similar spacetime dependence between Eq.~\eqref{eq:toy_RL} above and the expression in Eq.~\eqref{eq:GL_min_pot_EB}, which describes a contribution from the one-handed single $k$-mode in the minimal gravitational leptogenesis. In particular, the constant $E$ in our toy model resembles the (perturbative) parity-violating amplitude growth in Eq.~\eqref{eq:GL_minimal_mode}.
While the similarity in the spatial dependence (associated with the constant $B$) is somewhat subtle due to the additional assumption of $kz\ll 1$, the similarity in the electric component still gives us motivation to use our toy model~\eqref{eq:grav_tensor_pot} to investigate the chirality production in gravitational leptogenesis.

\bibliographystyle{JHEP}
\bibliography{ref}

\end{document}